\documentclass[atoms,article,accept,moreauthors,pdftex]{mdpi-arxiv} 

\firstpage{1} 
\pubvolume{1}
\issuenum{1}
\articlenumber{0}
\pubyear{2021}
\copyrightyear{2021}
\datereceived{\today} 
\dateaccepted{} 
\datepublished{} 
\hreflink{https://doi.org/} 

\usepackage{graphicx}
\usepackage{dcolumn}
\usepackage{bm}

\Title{Cartesian operator factorization method for Hydrogen}

\TitleCitation{Cartesian operator factorization method for Hydrogen}

\Author{Xinliang Lyu $^{1\dagger}$\orcidA{0000-0002-4077-5156}, Christina Daniel $^2$\orcidB{0000-0003-2680-435X}, and James K. Freericks$^2$*\orcidC{0000-0002-6232-9165}}

\AuthorNames{Xinliang Lyu, Christina Daniel, and James K. Freericks}

\AuthorCitation{Lyu, X.; Daniel, C.; Freericks, J. K.}

\address{%
$^{1}$ \quad College of Civil Engineering, Tongji University, 1239 Siping Road, Shanghai, China, 200092; lyu@issp.u-tokyo.ac.jp\\
$^{2}$ \quad Department of Physics, Georgetown University, 37th and O Sts. NW, Washington, DC 20057 USA;
cd1054@georgetown.edu and james.freericks@georgetown.edu}

\corres{Correspondence: james.freericks@georgetown.edu; Tel. +202-687-6159}

\firstnote{Current address: Institute for Solid State Physics, The University of Tokyo,  Kashiwa, Chiba 277-8581, Japan} 

\abstract{We generalize Schr\"odinger's factorization method for Hydrogen from the conventional separation into angular and radial coordinates to a Cartesian-based factorization. Unique to this approach, is the fact that the Hamiltonian is represented as a sum over factorizations in terms of coupled operators that depend on the coordinates and momenta in each Cartesian direction. We determine the eigenstates and energies, the wavefunctions in both coordinate and momentum space, and we also illustrate how this technique can be employed to develop the conventional confluent hypergeometric equation approach. The methodology developed here could potentially be employed for other Hamiltonians that can be represented as the sum over coupled Schr\"odinger factorizations.}


\begin{document}

\section{\label{sec: intro}Introduction}

The Hydrogen atom was originally solved by Pauli employing operator methods by discovering the Lie algebra of the SO(4) symmetry in the problem~\cite{pauli}. Schr\"odinger followed shortly with the differential equation approach of wave mechanics~\cite{schroedinger_wavea,schroedinger_waveb}. In 1940, Schr\"odinger developed the factorization method for quantum mechanics~\cite{schroedinger_factor1} and employed it to solve Hydrogen as well~\cite{schroedinger_factor2}. This factorization method was extensively reviewed by Infeld and Hull~\cite{infeld_hull}. Our focus in this work is on an operator-based method to solve Hydrogen. Similar to how the wave-equation approach can be solved directly in Cartesian coordinates~\cite{ajp_cartesian}, we develop here the Cartesian factorization method for Hydrogen. While it shares some of the characteristics of the spherical coordinate-based factorization method, it is distinctly different. It is surprising that one can discover a new methodology for solving Hydrogen nearly one hundred years since its first solution.

There are two new developments that arise with this solution. First, this approach generalizes the Schr\"odinger factorization method, which employs a single raising and lowering operator factorization, into an approach that works with the sum over three coupled raising and lowering operator factorizations---one for each Cartesian coordinate. These raising and lowering operators also depend on the radial coordinate, so the operators corresponding to the different Cartesian directions do not commute with each other. Nevertheless, a full operator-based approach can be employed to solve the quantum problem. Second, the strategy used here, where we solve first for the energy eigenstate as a sequence of operators acting on the ground state of an auxiliary Hamiltonian, allows us to then construct the eigenfunctions in both real space and momentum space by simply projecting onto the coordinate-space and momentum-space eigenfunctions, whereby both solutions proceed using the same methodology. This is quite different from conventional differential equation approaches, which are quite dissimilar for coordinate and momentum space. It does turn out that the algebra required for the momentum-space eigenfunctions is somewhat more complicated than for the real-space functions.

Conventional quantum mechanics suffers from overuse of the coordinate-space representation. This is solely for convenience---the Schr\"odinger equation is a second-order linear differential equation in coordinate space; in momentum space, it generically becomes an integral equation, which is much more difficult to work with. One might argue that this is just fine. After all, the Stone-von Neumann theorem~\cite{stone,von-neumann} tells us that all representations are equivalent to the coordinate-space representation. But, the ability to formulate quantum mechanics in a representation-independent fashion is an important cornerstone of the theory. In this work, we show how to determine wavefunctions independent of the basis.

One other notable result is that nearly all of the operator-based derivations we employ can be performed without using any calculus. We illustrate this throughout the paper. Calculus ends up being needed only for the normalization of the wavefunctions and for the derivation of some identities we require when calculating the momentum-space wavefunctions. Of course, one can map this problem onto a differential equation as well. We illustrate how one can extract the conventional confluent hypergeometric equation for the wavefunction of Hydrogen in coordinate space below.

While the differential equation approach to solving Hydrogen in spherical coordinates is performed in almost every quantum textbook, the operator method in spherical coordinates is limited to only a handful of textbooks. Starting with Harris and Loeb~\cite{harris_loeb}, and followed shortly thereafter by Green~\cite{green}, Bohm~\cite{bohm}, Ohanian~\cite{ohanian} and de Lange and Raab~\cite{delange_raab}, Schr\"odinger's factorization method (and the closely related ladder-operator method) has been covered amply in many texts. One might even say it has been increasing in popularity, as it has appeared in a number of recent texts as well, including Hecht~\cite{hecht}, Binney and Skinner~\cite{binney_skinner}, Schwabl~\cite{schwabl} and Razavy~\cite{razavy}. All of these texts employ essentially the same technique, closely related to Schr\"odinger's original method~\cite{schroedinger_factor2}. Only de Lange and Raab~\cite{delange_raab} also solve the problem in momentum space, but they employ a direct operator problem formulated in momentum space, which is a completely different technique and is not closely related to the coordinate-space approach. Judd~\cite{judd} also solves the problem directly in momentum space by determining the spherical harmonics of a four-sphere, following Fock's original derivation~\cite{fock}.

Here, we proceed differently. The Cartesian factorization of the Coulomb Hamiltonian was first worked out by Andrianov, Borisov and Ioffe in 1984~\cite{andrianov}. We begin by writing down the Hydrogen Hamiltonian $\hat{\mathcal H}$ as
\begin{equation}
    \hat{\mathcal H}=\frac{\hat p_x^2+\hat p_y^2+\hat p_z^2}{2m}-\frac{e^2}{\hat r},
    \label{eq: ham1}
\end{equation}
where we use hats to denote operators, $\hat p_\alpha$ ($\alpha=x,$ $y$, and $z$) are the momentum operators in the $\alpha$th direction, $\hat r_\alpha$ are the corresponding coordinate operators, and $[\hat r_\alpha,\hat p_\beta]=i\hbar\delta_{\alpha\beta}$, with $\hbar$ Planck's constant. We work in electrostatic units, where $|e|$ is the magnitude of the charge of a proton or electron, $m$ is the mass of the electron, and $\hat r^2=\hat r_x^2+\hat r_y^2+\hat r_z^2$. Reduced mass effects are taken into account by simply replacing $m\to m_em_H/(m_e+m_H)=\mu$, but all of our formulas here will use $m$ for the mass.

The Cartesian factorization rewrites the Hydrogen Hamiltonian in the following form:
\begin{equation}
    \hat{\mathcal H}=\sum_{\alpha=x,y,z}\hat A_\alpha^\dagger(\lambda{=}1)\hat A_\alpha(\lambda{=}1)+E(\lambda{=}1),
    \label{eq: ham2}
\end{equation}
where the raising and lowering operators are Hermitian conjugates $\hat A_\alpha^\dagger(\lambda)=[\hat A_\alpha(\lambda)]^\dagger$  given by
\begin{equation}
    \hat A_\alpha(\lambda)=\frac{1}{\sqrt{2m}}\left ( \hat p_\alpha-i\frac{\hbar}{\lambda a_0}\frac{\hat r_\alpha}{\hat r}\right ).
    \label{eq: lowering_def}
\end{equation}
Here, we have used $a_0=\hbar^2/me^2$, which is the Bohr radius, and that the energy satisfies
\begin{equation}
    E(\lambda)=-\frac{e^2}{2 a_0\lambda^2}.
    \label{eq: energy}
\end{equation}
In the Hydrogen Hamiltonian, we have $\lambda=1$, but we will be using these operators and energies for different values of $\lambda$ throughout the work. 

The verification that Eq.~(\ref{eq: ham2}) is equal to Eq.~(\ref{eq: ham1}) follows from a direct computation. We have
\begin{equation}
    \hat A_\alpha^\dagger(\lambda)\hat A_\alpha(\lambda)=\frac{\hat p_\alpha^2}{2m}-i\frac{\hbar}{2m\lambda a_0}\left [\hat p_\alpha,\frac{\hat r_\alpha}{\hat r}\right ]+\frac{\hbar^2}{2m\lambda^2 a_0^2}\frac{\hat r_\alpha^2}{\hat r^2}.
    \label{eq: com1}
\end{equation}
We use the fact that $[\hat r^2,\hat p_\alpha]=\sum_{\beta{=}x,y,z}[\hat r_\beta^2,\hat p_\alpha]=2i\hbar \hat r_\alpha$, which can be employed to show that $[\hat p_\alpha,1/\hat r]=i\hbar \hat r_\alpha/\hat r^3$ (for further details, see the appendix of Ref.~\cite{rushka} and Sec. II below). The commutator can then be computed, and yields
\begin{equation}
    \hat A_\alpha^\dagger(\lambda)\hat A_\alpha(\lambda)=\frac{\hat p^2_\alpha}{2m}
    -\frac{e^2}{2\lambda}\left ( \frac{1}{\hat r}-\frac{\hat r_\alpha^2}{\hat r^3}\right )+\frac{e^2}{2\lambda^2 a_0}\frac{\hat r_\alpha^2}{\hat r^2}.
    \label{eq: com2}
\end{equation}    
Summing over the spatial indices $\alpha$ immediately gives us
\begin{equation}
    \sum_{\alpha{=}x,y,z}\hat A_\alpha^\dagger(\lambda)\hat A_\alpha(\lambda)=\frac{\hat p_x^2+\hat p_y^2+\hat p_z^2}{2m}-\frac{e^2}{\lambda \hat r}+\frac{e^2}{2\lambda^2 a_0}.
    \label{eq: adag_a}
\end{equation}
Hence, we find that Eq.~(\ref{eq: ham2}) and Eq.~(\ref{eq: energy}) produce the Hydrogen Hamiltonian and ground state energy when we set $\lambda=1$.

Note that the raising and lowering operators do not commute with each other for different Cartesian coordinates. Hence, the factorization involves coupled operators and is different in character from
the Cartesian factorization of the isotropic simple harmonic oscillator in three dimensions, where the raising and lowering operators for different directions commute with each other.

The form of the Hamiltonian in Eq.~(\ref{eq: ham2}) allows us to directly determine the ground-state energy and the ground-state wavefunction in coordinate space. Because the factorized form of the Hamiltonian is the sum of positive semidefinite operator terms and a constant, the ground-state energy is given by the constant $E_{\rm gs}=E(\lambda{=}1)=-e^2/2a_0$ and the ground-state wavefunction satisfies
\begin{equation}
    \hat A_\alpha(\lambda{=}1)|\phi_{\lambda{=}1}\rangle=0,
    \label{eq: ground_state}
\end{equation}
for all $\alpha=x,y,z$. We use the symbol $|\phi_\lambda\rangle$ to denote the ground state of the auxiliary Hamiltonian
\begin{equation}
    \hat{\mathcal H}(\lambda)=\sum_{\alpha{=}x,y,z}\hat A_\alpha^\dagger(\lambda)\hat A_\alpha(\lambda)+E(\lambda)=\frac{\hat p_x^2+\hat p_y^2+\hat p_z^2}{2m}-\frac{e^2}{\lambda \hat r},
    \label{eq: ham_aux}
\end{equation}
which satisfies
\begin{equation}
    \hat A_\alpha(\lambda)|\phi_\lambda\rangle=0
\end{equation}
for fixed $\lambda$ (and all $\alpha=x,y,z$) with energy $E(\lambda)$.

We find the ground-state wavefunction $\psi_{gs}(r_x,r_y,r_z)$ by taking the overlap of Eq.~(\ref{eq: ground_state}) with the position eigenstate bra $\langle r_x,r_y,r_z|$:
\begin{equation}
    \psi_{gs}(r_x,r_y,r_z)=\langle r_x,r_y,r_z|\phi_{\lambda{=}1}\rangle.
\end{equation}
Using the condition for the ground-state wavefunction in Eq.~(\ref{eq: ground_state}), we find that
\begin{equation}
    \langle r_x,r_y,r_z|\hat p_\alpha|\phi_{\lambda{=}1}\rangle=i\frac{\hbar}{a_0}\frac{r_\alpha}{r}\psi_{gs}(r_x,r_y,r_z).
    \label{eq: gs_wf1}
\end{equation}
We can proceed in one of two ways at this point. We can use the coordinate-space representation of the momentum operator, given by $\hat p_\alpha=-i\hbar \partial/\partial r_\alpha$, to
find the three differential equations
\begin{equation}
    \frac{\partial}{\partial r_\alpha}\psi_{gs}(r_x,r_y,r_z)=-\frac{r_\alpha}{a_0 r}\psi_{gs}(r_x,r_y,r_z).
    \label{eq: gs_diffeq}
\end{equation}
These three equations can be immediately solved via
\begin{equation}
    \psi_{gs}(r_x,r_y,r_z)=\psi_{gs}(r_x{=}0,r_y{=}0,r_z{=}0)e^{-\frac{r}{a_0}}.
    \label{eq: gs_final}
\end{equation}
This is the well-known ground-state solution for Hydrogen, with the overall normalization constant still needing to be determined.

We can also solve this problem without calculus. But it turns out the algebra for this is a bit more complex. We illustrate why here and provide some additional details in the appendix. First, we use the translation operator to write
\begin{equation}
    |r_x,r_y,r_z\rangle=e^{-\frac{i}{\hbar}(r_x\hat p_x+r_y\hat p_y+r_z\hat p_z)}|r_x{=}0,r_y{=}0,r_z{=}0\rangle.
    \label{eq: translation}
\end{equation}
Then we compute
\begin{equation}
    \psi_{gs}(r_x,r_y,r_z)=\langle r_x{=}0,r_y{=}0,r_z{=}0|\left [\sum_{n{=}0}^\infty \frac{1}{n!} \left (\sum_{\alpha{=}x,y,z}\frac{ir_\alpha\hat p_\alpha}{\hbar}\right )^n\right ]|\phi_{\lambda{=}1}\rangle.
    \label{eq: gs_wf}
\end{equation}
To evaluate this expression, we want to use Eq.~(\ref{eq: ground_state}) to replace $\hat p_\alpha$ by $i\hbar\hat r_\alpha/(a_0\hat r)$. There are two issues that arise for this: (i) one cannot immediately evaluate operators like $\hat r_\alpha/\hat r$ onto the coordinate eigenstate at the origin, without some appropriate limiting procedure and (ii) the fact that $\hat p_\alpha$ does not commute with $\hat r_\alpha/\hat r$ creates many additional terms that one must carefully keep track of. It is easier to discuss a strategy for how to work on this after one has developed a bit more formalism, so this discussion continues in the Appendix. The full expression of the translation operator and the methodology needed to complete the algebraic determination of the maximal angular momentum wavefunctions has already been completed elsewhere~\cite{rushka}. 

The momentum wavefunction is also not so simple to determine and we postpone discussing it until we work on the general case below.

The organization of the remainder of the paper is as follows: in Sec. II, we derive the operator form for the eigenfunctions and eigenvalues for the general case after briefly summarizing the properties of harmonic polynomials. The coordinate-space wavefunctions are derived in Sec. III and the momentum space wavefunctions in Sec. IV. Section V provides a derivation of the more conventional confluent hypergeometric equation approach. We conclude in Sec. VI.

\section{Deriving the Energy Eigenstate}

We begin with a short ``tutorial'' on how to compute commutators without taking derivatives or working with an explicit representation for the operators. This methodology was introduced by Dirac~\cite{dirac}, and is an elegant way to determine commutators without calculus. A rather complete discussion can also be found in the appendix of Ref.~\cite{rushka}. We employ this methodology throughout.

To begin, we note that the commutator of $\hat r^2=\hat r_x^2+\hat r_y^2+\hat r_z^2$ with $\hat p_\alpha$, is simple to compute, but by also employing the product rule, we quickly learn that
\begin{equation}
    [\hat r^2,\hat p_\alpha]=2i\hbar \hat r_\alpha=\hat r[\hat r,\hat p_\alpha]+[\hat r,\hat p_\alpha]\hat r,
\end{equation}
or $[\hat r,\hat p_\alpha]=i\hbar\hat r_\alpha/\hat r$. Note that we have to move the commutator through $\hat{r}$ to complete this derivation. This is easy to establish \textit{a posteriori}, but it can actually be directly verified by the Jacobi identity and the fact that the operator $\hat{r}^2$ has the same eigenvectors as the operator $\hat{r}$~\cite{bohm,rushka}. With a simple inductive argument, one can then establish that $[\hat r^n,\hat p_\alpha]=i\hbar n\hat r_\alpha \hat r^{n-2}$, for all integers $n$.

Next, we work with the Hermitian (but not self-adjoint) radial momentum operator, $\hat p_r$, defined to be
\begin{equation}
    \hat p_r=\frac{1}{\hat r}\sum_{\alpha{=}x,y,z}\hat r_\alpha\hat p_\alpha-i\frac{\hbar}{\hat r}.
    \label{eq: rad_mom_def}
\end{equation}
One immediately sees that
\begin{equation}
    [\hat r,\hat p_r]=\left [ \hat r,\frac{1}{\hat r}(\hat r_x\hat p_x+\hat r_y\hat p_y+\hat r_z\hat p_z)-i\frac{\hbar}{\hat r}\right ]=i\hbar,
    \label{eq: rad_commute}
\end{equation}
after using the identity derived above and the fact that $\hat r_x^2+\hat r_y^2+\hat r_z^2=\hat r^2$. The radial momentum determines the radial component of the kinetic energy $\hat T$, separating it into the radial component and the perpendicular component $\hat T_\perp$ via
\begin{equation}
    \hat T_\perp=\hat T-\frac{\hat p_r^2}{2m}=\frac{\hat p_x^2+\hat p_y^2+\hat p_z^2}{2m}-\frac{\hat p_r^2}{2m}.
\end{equation}

The perpendicular component of the kinetic energy commutes with the radial coordinate, as can be seen by direct computation:
\begin{align}
    [\hat r,\hat T_\perp]&=\frac{1}{2m}[\hat r,\hat p_x^2+\hat p_y^2+\hat p_z^2-\hat p_r^2]\nonumber\\
    &=\frac{i\hbar}{2m} \left (\hat p_x \frac{\hat r_x}{\hat r}+ \frac{\hat r_x}{\hat r}\hat p_x+\hat p_y \frac{\hat r_y}{\hat r}+ \frac{\hat r_y}{\hat r}\hat p_y+\hat p_z \frac{\hat r_z}{\hat r}+ \frac{\hat r_z}{\hat r}\hat p_z-2\hat p_r\right )=0.
    \label{eq: tperp_r_com}
\end{align}
The cancellation follows from a simple rearrangement of operators using the commutation relation and the definition of the radial momentum in Eq.~(\ref{eq: rad_mom_def}).

One of the subtleties we have to work with is that the Cartesian components of momentum do not commute with the radial momentum. In particular, we find that
\begin{align}
[\hat p_r,\hat p_\alpha]&=\left [ \frac{1}{\hat r}(\hat r_x\hat p_x+\hat r_y\hat p_y+\hat r_z\hat p_z-i\hbar),\hat p_\alpha\right ]\nonumber\\
&=\frac{1}{\hat r}i\hbar\hat p_\alpha-i\hbar \frac{\hat r_\alpha}{\hat r^3}(\hat r_x\hat p_x+\hat r_y\hat p_y+\hat r_z\hat p_z-i\hbar)\nonumber\\
&=i\hbar\frac{1}{\hat r}\hat p_\alpha-i\hbar\frac{\hat r_\alpha}{\hat r^2}\hat p_r.
\end{align}
Similarly, the Cartesian components of position do not commute with the radial momentum
\begin{equation}
[\hat r_\alpha,\hat p_r]=\left [\hat r_\alpha, \frac{1}{\hat r}(\hat r_x\hat p_x+\hat r_y\hat p_y+\hat r_z\hat p_z-i\hbar)\right ]=i\hbar\frac{\hat r_\alpha}{\hat r}.
\end{equation}

These results allow us to compute our final commutator, that of the radial momentum with the perpendicular kinetic energy:
\begin{align}
[\hat p_r,\hat T_\perp]&=\frac{1}{2m}(
\hat p_x[\hat p_r,\hat p_x]+[\hat p_r,\hat p_x]\hat p_x+\hat p_y[\hat p_r,\hat p_y]+[\hat p_r,\hat p_y]\hat p_y+\hat p_z[\hat p_r,\hat p_z]+[\hat p_r,\hat p_z]\hat p_z )\nonumber\\
&=2i\hbar\frac{1}{\hat r}\hat T_\perp,
\label{eq: tperp_pr_com}
\end{align}
which follows after some complex algebra.

We do not use a separation of variables into radial and angular coordinates, nevertheless, similar to the Cartesian differential equation approach~\cite{ajp_cartesian}, we need to work with harmonic polynomials, but here in operator form. Kramers originated the use of harmonic polynomials for angular momentum~\cite{kramers} (see also Brinkman~\cite{brinkman} and Powell and Crasemann~\cite{powell_crasemann}), which has been revitalized recently by Weinberg~\cite{weinberg}. Here, we follow the approach of Avery~\cite{avery} and define the $l$th-order harmonic polynomial $P_h^l(\hat r_x,\hat r_y,\hat r_z)$ to be a homogeneous polynomial, so that
\begin{equation}
    \left [ \hat r \hat p_r,P_h^l(\hat r_x,\hat r_y,\hat r_z)\right ]=-i\hbar l P_h^l(\hat r_x,\hat r_y,\hat r_z),
    \label{eq: p_harm_homog}
\end{equation}
for $l$ a nonnegative integer.
Here, we have $\hat r\hat p_r=\hat r_x\hat p_x+\hat r_y\hat p_y+\hat r_z\hat p_z-i\hbar$.
Note that the radial momentum operator is a Hermitian operator that is conjugate to the radial coordinate operator and satisfies Eq.~(\ref{eq: rad_commute}).

The result in Eq.~(\ref{eq: p_harm_homog}) follows by simply evaluating the commutator, which yields a term given by $-i\hbar$ times the monomial term for every coordinate factor in each monomial term of the polynomial; when the polynomial is homogeneous, the number of terms in each monomial is the same, and hence the entire polynomial is multiplied by $-i\hbar l$. In addition, these polynomials satisfy Laplace's equation, written in the operator form
\begin{equation}
    \sum_{\alpha{=}x,y,z}\left [ \hat p_\alpha, \left [ \hat p_\alpha, P_h^l(\hat r_x,\hat r_y,\hat r_z)\right ] \right ]=0.
    \label{eq: laplace}
\end{equation}
These harmonic polynomials are defined up to an overall multiplicative factor. The low-order ones are just what we expect them to be from our knowledge of Cartesian spherical harmonics. The zeroth-order one is just $1$. The first-order ones are $\hat r_x$, $\hat r_y$, and $\hat r_z$. The second-order ones are $\hat r_x\hat r_y$, $\hat r_y\hat r_z$, $\hat r_z\hat r_x$, $\hat r_x^2-\hat r_y^2$, and $-\hat r_x^2-\hat r_y^2+2\hat r_z^2$. One should note that the harmonic polynomials have all the $\hat r^2$ dependence removed from them, because the Laplacian of $\hat r^2$ is nonzero. We will be employing the harmonic polynomial operators heavily in the derivation below.

But, before getting there, we want to note one other special property about the harmonic polynomials. First, we can easily verify from the commutation relations derived above that
\begin{equation}
\left [ \hat p_r,\frac{\hat r_\alpha}{\hat r}\right ]=0,
\end{equation}
which then implies that
\begin{equation}
    \left [\hat p_r,\frac{1}{\hat r^l}P_h^l(\hat r_x,\hat r_y,\hat r_z)\right ]=0.
    \label{eq: rad_mom_commutes}
\end{equation}
The factor $1/\hat r^l$ is needed to divide each of the $l$ $\hat r_\alpha$ factors in each monomial term of the harmonic polynomial to ensure that each monomial commutes with the radial momentum.

Now with these technical details finished, we are ready to move on to the derivation of the eigenfunctions and eigenvalues of Hydrogen. In Schr\"odinger's original factorization method, he constructed higher-energy states by acting a chain of operators on the ground state of an auxiliary Hamiltonian, which was related to the original Hamiltonian, but employed a series of different raising and lowering operators, as well as different constant terms. He derived this result, by showing that the wavefunction constructed must be an eigenfunction of the original Hamiltonian. We will follow  similar strategy here, but it has a few places where the steps are modified because our Hamiltonian has a sum of factorized terms instead of just one. While it may seem like we are just going to guess the solution and then verify it, which is a valid approach, the motivation for the guess comes from the standard operator formalism solution for Hydrogen.

To begin, we first must define another set of raising and lowering operators that we call $\hat B^\dagger_r(\lambda)$ and $\hat B_r(\lambda)$, with $\hat B_r^\dagger(\lambda)=[\hat B_r(\lambda)]^\dagger$. The operator is defined by
\begin{equation}
    \hat B_r(\lambda)=\frac{1}{\sqrt{2m}}\left [ \hat p_r-i\hbar\left (\frac{1}{(\lambda +1)a_0}-\frac{\lambda +1}{\hat r}\right )\right ],
    \label{eq: b_def}
\end{equation}
following de Lange and Raab's convention~\cite{delange_raab}. This operator can be expressed in terms of our Cartesian operators as follows:
\begin{equation}
    \hat B_r(\lambda)=\sum_{\alpha{=}x,y,z}\frac{\hat r_\alpha}{\hat r}\hat A_\alpha(\lambda+1)+i\frac{\lambda\hbar}{\sqrt{2m}\hat r}.
\end{equation}
One can immediately verify that
\begin{align}
    \hat{\mathcal H}(\lambda{=}1)&=\hat B_r^\dagger(\lambda)\hat B_r(\lambda)+\hat T_\perp-\frac{\hbar^2\lambda(\lambda+1)}{2m\hat r^2}+E(\lambda +1)\nonumber\\
    &=\hat B_r(\lambda)\hat B_r^\dagger(\lambda)+\hat T_\perp-\frac{\hbar^2(\lambda+1)(\lambda+2)}{2m\hat r^2}+E(\lambda +1).
    \label{eq: br_intertwine1}
\end{align}

Before we construct the ansatz for the eigenstate, we need to establish the following identity:
\begin{equation}
    \hat T_\perp P_h^l(\hat r_x,\hat r_y,\hat r_z)|\phi_{\lambda}\rangle=\frac{\hbar^2 l(l+1)}{2m\hat r^2}P_h^l(\hat r_x,\hat r_y,\hat r_y)|\phi_{\lambda}\rangle.
    \label{eq: tperp_identity}
\end{equation}
The proof requires a number of steps. First, we establish the action of the radial momentum on the auxiliary Hamiltonian ground state via
\begin{align}
    \hat p_r|\phi_{\lambda}\rangle&=\left (\sum_\alpha \frac{\hat r_\alpha}{\hat r}\hat p_\alpha-\frac{i\hbar}{\hat r}\right )|\phi_\lambda\rangle=\left (\sum_\alpha \frac{\hat r_\alpha}{\hat r}\frac{i\hbar\hat r_\alpha}{\lambda a_0\hat r}-\frac{i\hbar}{\hat r}\right )|\phi_\lambda\rangle\nonumber\\
    &=i\hbar\left (\frac{1}{\lambda a_0}-\frac{1}{\hat r}\right )|\phi_\lambda\rangle,
    \label{eq: pr_state}
\end{align}
where we used the fact that $\hat A_\alpha(\lambda)|\phi_\lambda\rangle=0$ implies that
\begin{equation}
    \hat p_\alpha |\phi_\lambda\rangle=\frac{i\hbar\hat r_\alpha}{\lambda a_0\hat r}|\phi_\lambda\rangle
\end{equation}
for all $\alpha=x,y,z$. Second, we show that $\hat T_\perp|\phi_\lambda\rangle=0$.
This is done with the following steps:
\begin{align}
\hat T_\perp|\phi_\lambda\rangle&=\left (\hat T-\frac{\hat p_r^2}{2m}\right )|\phi_\lambda\rangle=\left (\hat{\mathcal H}(\lambda)+\frac{e^2}{\lambda\hat r}-\frac{\hat p_r^2}{2m}\right )|\phi_\lambda\rangle\nonumber\\
&=\left (-\frac{e^2}{2\lambda^2 a_0}+\frac{e^2}{\lambda\hat r}-i\frac{\hbar}{2m}\hat p_r\left [\frac{1}{\lambda a_0}-\frac{1}{\hat r}\right ]\right )|\phi_\lambda\rangle\nonumber\\
&=\left (-\frac{e^2}{2\lambda^2 a_0}+\frac{e^2}{\lambda\hat r}-\frac{\hbar^2}{2m\hat r^2}+\frac{\hbar^2}{2m}\left [\frac{1}{\lambda a_0}-\frac{1}{\hat r}\right ]^2\right )|\phi_\lambda\rangle\nonumber\\
&=0,
\end{align}
where Eq.~(\ref{eq: pr_state}) is applied twice.
Next, we evaluate $\hat T_\perp$ acting on the harmonic polynomial and the auxiliary Hamiltonian ground-state via
\begin{equation}
\hat T_\perp P_h^l(\hat r_x,\hat r_y,\hat r_z)|\phi_{\lambda}\rangle=\left [\hat T_\perp, P_h^l(\hat r_x,\hat r_y,\hat r_z)\right ]|\phi_{\lambda}\rangle=\left [\hat T-\frac{\hat p_r^2}{2m}, P_h^l(\hat r_x,\hat r_y,\hat r_z)\right ]|\phi_{\lambda}\rangle
\end{equation}
because $\hat T_\perp|\phi_{\lambda}\rangle=0$.
We evaluate each piece of the commutator next. First we compute
\begin{align}
[\hat T,P_h^l(\hat r_x,\hat r_y,\hat r_z)]|\phi_{\lambda}\rangle&=\frac{1}{2m}\sum_\alpha\left ( \hat p_\alpha [\hat p_\alpha,P_h^l(\hat r_x,\hat r_y,\hat r_z)]+[\hat p_\alpha,P_h^l(\hat r_x,\hat r_y,\hat r_z)]\hat p_\alpha\right )|\phi_{\lambda}\rangle\nonumber\\
&=\frac{1}{2m}\sum_\alpha\left ([\hat p_\alpha,[\hat p_\alpha,P_h^l(\hat r_x,\hat r_y,\hat r_z)]] +2[\hat p_\alpha,P_h^l(\hat r_x,\hat r_y,\hat r_z)]\frac{i\hbar\hat r_\alpha}{\lambda a_0\hat r}\right )|\phi_{\lambda}\rangle\nonumber\\
&=\frac{e^2l}{\lambda\hat r}P_h^l(\hat r_x,\hat r_y,\hat r_z)|\phi_{\lambda}\rangle.
\label{eq: tperp_com1}
\end{align}
In the second line, the double commutator vanishes because the harmonic polynomial satisfies Laplace's equation as shown in Eq.~(\ref{eq: laplace}), while the second term simplifies because the harmonic polynomial is homogeneous, as shown in Eq.~(\ref{eq:  p_harm_homog}). We used the definition of the Bohr radius to simplify the final result as well.

The second piece of the commutator is
\begin{align}
-\left [ \frac{\hat p_r^2}{2m},P_h^l(\hat r_x,\hat r_y,\hat r_z)\right ]|\phi_{\lambda}\rangle&=-\left [ \frac{\hat p_r^2}{2m},\frac{\hat r^lP_h^l(\hat r_x,\hat r_y,\hat r_z)}{\hat r^l}\right ]|\phi_{\lambda}\rangle=-\frac{P_h^l(\hat r_x,\hat r_y,\hat r_z)}{ 2m\hat r^l}[\hat p_r^2,\hat r^l]|\phi_{\lambda}\rangle\nonumber\\
&=-\frac{P_h^l(\hat r_x,\hat r_y,\hat r_z)}{ 2m\hat r^l}i\hbar (\hat p_r l\hat r^{l-1}+l\hat r^{l-1}\hat p_r)|\phi_{\lambda}\rangle\nonumber\\
&=-\frac{\hbar^2P_h^l(\hat r_x,\hat r_y,\hat r_z)}{ 2m\hat r^l}\left (2l\left [\frac{1}{\lambda a_0}-\frac{1}{\hat r}\right ]\hat r^{l-1}-l(l-1)\hat r^{l-2}\right )|\phi_{\lambda}\rangle\nonumber\\
&=P_h^l(\hat r_x,\hat r_y,\hat r_z)\left (\frac{\hbar^2 l(l+1)}{2m\hat r^2}-\frac{e^2l}{\lambda\hat r}\right )|\phi_{\lambda}\rangle.
\label{eq: tperp_com2}
\end{align}
Adding Eqs.~(\ref{eq: tperp_com1}) and (\ref{eq: tperp_com2}) establishes the identity in Eq.~(\ref{eq: tperp_identity}).

We also need to determine how $\hat T_\perp-\hbar^2l(l+1)/(2m\hat r^2)$ commutes 
with $\hat B_r^\dagger(\lambda)$. A direct computation, employing Eqs.~(\ref{eq: tperp_r_com}) and (\ref{eq: tperp_pr_com}) yields
\begin{equation}
    \left [\hat T_\perp-\frac{\hbar^2 l(l+1)}{2m\hat r^2},\hat B_r^\dagger(\lambda)\right ]=-\frac{2i\hbar}{\sqrt{2m}\hat r}\left (\hat T_\perp-\frac{\hbar^2 l(l+1)}{2m\hat r^2}\right ).
    \label{eq: tperp_bdag_com}
\end{equation}
This equation can be rearranged to an intertwining relation
\begin{equation}
\left [\hat T_\perp-\frac{\hbar^2 l(l+1)}{2m\hat r^2}\right ]\hat B_r^\dagger(\lambda)=\left [\hat B_r^\dagger(\lambda)-
\frac{2i\hbar}{\sqrt{2m}\hat r}\right ]\left [\hat T_\perp-\frac{\hbar^2 l(l+1)}{2m\hat r^2}\right ]
\label{eq: br_tperp_interchange}
\end{equation}
indicating that the operator $\hat T_\perp-\hbar^2l(l+1)/(2m\hat r^2)$ can be moved through $\hat B_r^\dagger(\lambda)$ by shifting the latter operator to $\hat B_r^\dagger(\lambda)-2i\hbar/(\sqrt{2m}\hat r)$.
One can derive a similar intertwining relation given by
\begin{equation}
\frac{\hbar^2}{2m\hat r^2}\hat B_r^\dagger(\lambda)=\left [\hat B_r^\dagger(\lambda)-
\frac{2i\hbar}{\sqrt{2m}\hat r}\right ]\frac{\hbar^2}{2m\hat r^2}.
\label{eq: r_tperp_interchange}
\end{equation}

In the Schr\"odinger factorization method, an intertwining relation is employed to construct the higher-energy eigenstates of the original Hamiltonian from a series of raising operators acting on an auxiliary Hamiltonian ground state. Here, we employ the exact same technique, except the intertwining relation is complicated due to the fact that it generates many extra terms proportional to $\hat T_\perp-\hbar^2l(l+1)/(2m\hat r^2)$. Since these terms vanish when they operate on $P_h^l(\hat r_x,\hat r_y,\hat r_z)|\phi_{\lambda}\rangle$, they do not corrupt the final intertwining relation, but they make the derivation more cumbersome. We now go carefully through the construction of all of the remaining eigenstates.

We start with the modification of the intertwining relation, which is derived by commuting $\hat{\mathcal H}(1)$ through $\hat B_r^\dagger(\lambda)$:
\begin{align}
\hat{\mathcal H}(1)\hat B_r^\dagger(\lambda)&=
\Biggr [\hat B_r^\dagger(\lambda)\hat B_r(\lambda)+\hat T_\perp-\frac{\hbar^2\lambda(\lambda+1)}{2m\hat r^2}+E(\lambda+1)\Biggr ]\hat B_r^\dagger(\lambda)\nonumber\\
&=\hat B_r^\dagger(\lambda)\Biggr [\hat B_r(\lambda)\hat B_r^\dagger(\lambda)+\hat T_\perp-\frac{\hbar^2\lambda(\lambda+1)}{2m\hat r^2}+E(\lambda+1)\Biggr ]\nonumber\\
&~~~~~~-\frac{2i\hbar}{\sqrt{2m}\hat r}\left (\hat T_\perp-\frac{\hbar^2\lambda(\lambda+1)}{2m\hat r^2}\right )\nonumber\\
&=\hat B_r^\dagger(\lambda)\Biggr [ \hat{\mathcal H}(1)+\frac{\hbar^2 (\lambda+1)}{m\hat r^2}\Biggr ]-\frac{2i\hbar}{\sqrt{2m}\hat r}\left (\hat T_\perp-\frac{\hbar^2\lambda(\lambda+1)}{2m\hat r^2}\right ).
\label{eq: br_intertwine2}
\end{align}
This derivation proceeds by first using the top relation in Eq.~(\ref{eq: br_intertwine1}). Next, we construct the Hamiltonian via the lower relation in Eq.~(\ref{eq: br_intertwine1}) and also employ the commutator in Eq.~(\ref{eq: tperp_bdag_com}). This differs from the conventional intertwining relation due to the extra term on the last line of Eq.~(\ref{eq: br_intertwine2}). 

We now show how to determine the energy eigenstates using a quasi-inductive argument. We build up subsequent eigenstates and see why, in each case, the $\lambda$ values used are integers chosen according to a specific procedure. This ends up being a constructive methodology for generating the energy eigenstates.

We begin with an ansatz for the first (unnormalized) eigenstate in the chain of eigenstates. It has the form (with the subscript $nn-1$ of $\psi$ to be understood {\it a posteriori})
\begin{equation}
|\psi_{nn-1}\rangle=\hat r^{\alpha-n+1}P_h^{n-1}(\hat r_x,\hat r_y,\hat r_z)|\phi_\lambda\rangle.
\end{equation}
Here, the parameter $\lambda$ on the right hand side is a parameter, and our
initial goal is to show that we must have $\lambda=n$ and $\alpha=n-1$. To do so, we act ${\mathcal H}(\lambda{=}1)$ onto this state, which yields
\begin{equation}
{\mathcal H}(1)|\psi_{nn-1}\rangle=\left ( \frac{\hat p_r^2}{2m}+\hat T_\perp -\frac{e^2}{\hat r}\right )\hat r^{\alpha-n+1} P_h^{n-1}(\hat r_x,\hat r_y,\hat r_z) |\phi_\lambda\rangle.
\label{eq: first_in_chain}
\end{equation}
Our first step is to move $\hat{T}_\perp$ through the powers of $\hat{r}$. This is possible because $[\hat T_\perp,\hat r]=0$, as we now show
\begin{align}
[\hat T_\perp,\hat r]&=\frac{1}{2m}[\hat p_x^2+\hat p_y^2+\hat p_z^2-\hat p_r^2,\hat r]=\frac{1}{2m}\left (\sum_{\alpha=x,y,z}\left \{ \hat p_\alpha [\hat p_\alpha,\hat r]+[\hat p_\alpha,\hat r]\hat p_\alpha\right \}+2i\hbar\hat p_r\right )\nonumber\\
&=-\frac{i\hbar}{2m}\sum_{\alpha=x,y,z} \left (\hat p_\alpha \frac{\hat r_\alpha}{\hat r}+\frac{\hat r_\alpha}{\hat r}\hat p_\alpha\right )+\frac{i\hbar\hat p_r}{m}=0.
\end{align}
Verifying this result required us to use three facts: $[\hat r,\hat p_\alpha]=i\hbar\hat r_\alpha/\hat r$; $[\hat r,\hat p_r]=i\hbar$ and the definition of the radial momentum in Eq.~(\ref{eq: rad_mom_def}).

This commutator means we can move the $\hat{T}_\perp$ operator through the power of $\hat{r}$ and have it act on the harmonic polynomial operator and the state. We use the identity in Eq.~(\ref{eq: tperp_identity}) to replace the $\hat T_\perp$ term by a number, yielding
\begin{equation}
\hat{\mathcal H}(1)|\psi_{nn-1}\rangle=\left ( \frac{\hat p_r^2}{2m}+\frac{\hbar^2 n(n-1)}{2m\hat r^2} -\frac{e^2}{\hat r}\right )\hat r^{\alpha-n+1} P_h^{n-1}(\hat r_x,\hat r_y,\hat r_z) |\phi_\lambda\rangle.
\label{eq: eigenstate1}
\end{equation}
What remains to do, is operating $\hat p_r^2$ onto the state. We need Eq.~(\ref{eq: pr_state}) for the action of the radial momentum onto the auxiliary Hamiltonian ground state and the fact that we can commute $P_h^{n-1}(\hat r_x,\hat r_y,\hat r_z)/\hat r^{n-1}$ through $\hat p_r$. This leaves us with the final calculation
\begin{align}
\hat p_r^2\hat r^\alpha|\phi_\lambda\rangle&=\hat p_r\left ( [\hat p_r,\hat r^\alpha]+\hat r^\alpha \hat p_r\right )|\phi_\lambda\rangle=i\hbar\hat p_r\left ( -(\alpha+1)\hat r^{\alpha-1}+\frac{\hat r^\alpha}{\lambda a_0}\right )|\phi_\lambda\rangle\\
&=(i\hbar)^2\left ( \alpha(\alpha+1)\hat r^{\alpha-2}-2\frac{\alpha+1}{\lambda a_0}\hat r^{\alpha-1}+\frac{\hat r^\alpha}{\lambda^2 a_0^2}\right )|\phi_\lambda\rangle.\nonumber
\end{align}
Substituting into Eq.~(\ref{eq: eigenstate1}) and using the definition of the Bohr radius gives us
\begin{align}
\hat{\mathcal H}(1)|\psi_{nn-1}\rangle&=\left (\frac{\hbar^2[n(n-1)-\alpha(\alpha+1)]}{2m\hat r^2}+\frac{e^2}{\lambda\hat r}(\alpha+1-\lambda)-\frac{e^2}{2\lambda^2 a_0}\right )\nonumber\\&~~~~~~~~~~~~~~~\times\hat r^{\alpha-n+1}P_h^{n-1}(\hat r_x,\hat r_y,\hat r_z)|\phi_\lambda\rangle\nonumber\\
&=-\frac{e^2}{2n^2 a_0}|\psi_{nn-1}\rangle.
\end{align}
Clearly, we must have $\alpha=\lambda-1$ and $\alpha(\alpha+1)=n(n-1)$ (which means $\lambda=n$, because we also require $\lambda\ge 0$) in order for this to be an eigenstate, with the eigenvalue given on the last line. This establishes the correct integers for the first eigenstate in the series. The requirement that $\lambda\ge 0$ arises when we construct the ground-state wavefunction (see the Appendix)---if we do not make this requirement, then the ground-state wavefunction is not normalizable. We apply this constraint here, knowing this requirement.

\begin{figure}
    \centering
    \includegraphics[width=0.65\textwidth]{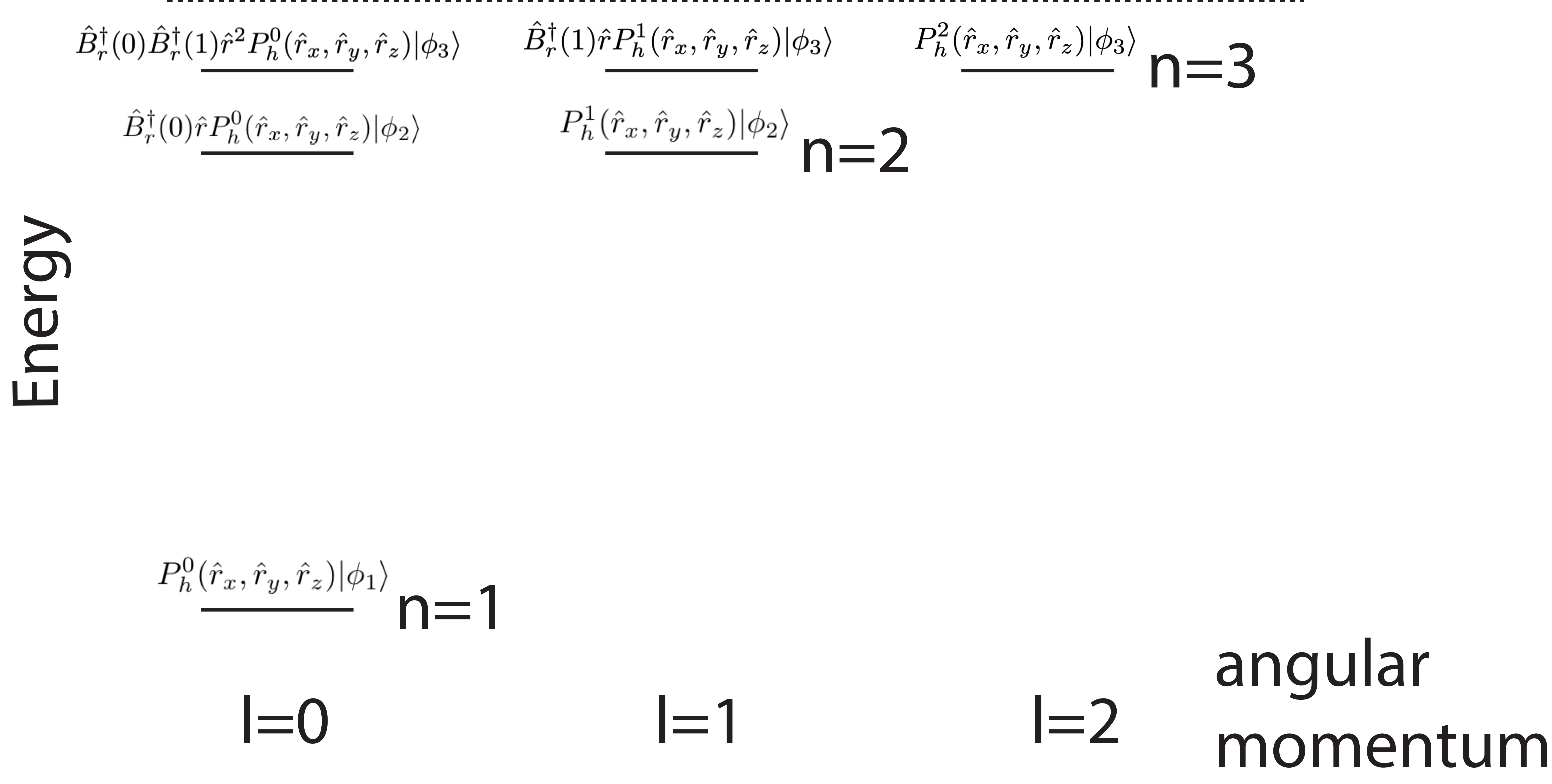}
    \caption{First three energy eigenstate multiplets, plotted to show the energy and the corresponding state, along with the angular momentum. The energies are the same for all inequivalent harmonic polynomials with the same $l$ value. The dotted line is the limit point for the energy eigenvalues.}
    \label{fig:energy}
\end{figure}

\begin{figure}
    \centering
    \includegraphics[width=0.75\textwidth]{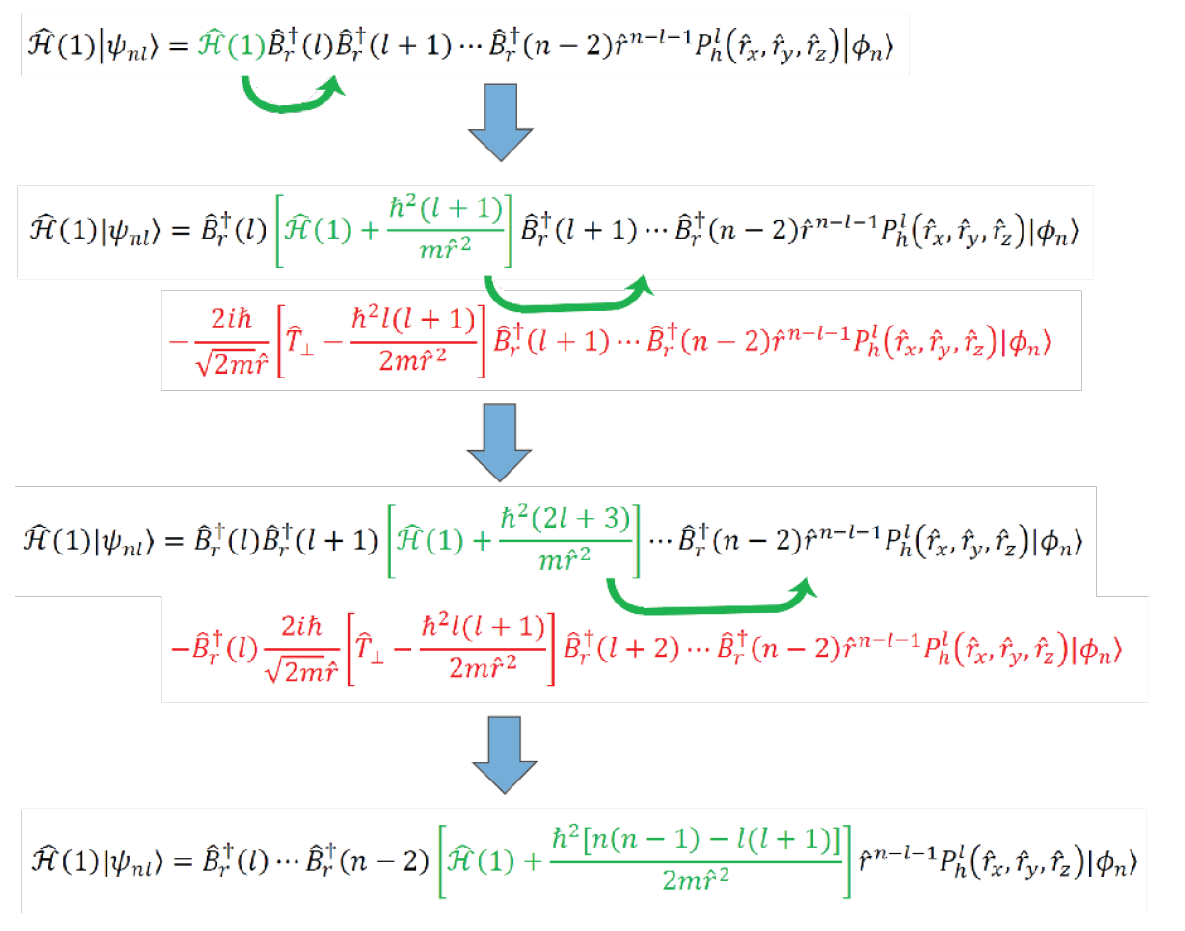}
    \caption{Schematic for how the energy eigenstate calculation works.}
    \label{fig:schematic}
\end{figure}

For the general case, we choose a set of real and nonnegative parameters $\{\nu_l,\cdots,\nu_{n-2}\}$ and form the trial eigenstate via
\begin{equation}
|\psi_{nl}\rangle=\hat B_r^\dagger(\nu_l)\hat B_r^\dagger(\nu_{l+1})\cdots\hat B_r^\dagger(\nu_{n-2})\hat r^{\alpha-l}P_h^l(\hat r_x,\hat r_y,\hat r_z)|\phi_\lambda\rangle.
\end{equation}
When we act ${\mathcal H}(1)$ from the left, we can move it through the terms one by one to the right. It will generate a number of terms proportional to $\hat T_\perp-\hbar^2C/(2m\hat r^2)$, for some number $C$, but those will all vanish when operated to the right against the state, because we will organize the terms so that $C=l(l+1)$. We are then left with
\begin{equation}
\hat{\mathcal H}(1)|\psi_{nl}\rangle=\hat B_r^\dagger(\nu_l)\cdots \hat B_r^\dagger(\nu_{n-2})\left ({\mathcal H}(1)+\frac{\hbar^2}{m\hat r^2}\sum_{j=l}^{n-2}(\nu_j+1)\right )\hat r^{\alpha-l} P_h^l(\hat r_x,r_y,r_z)|\phi_\lambda\rangle.
\end{equation} 
Acting $\hat{\mathcal H}(1)$ to the right proceeds exactly like what we did above and yields
\begin{align}
\hat{\mathcal H}(1)\hat r^{\alpha-l}P_h^l(\hat r_x,\hat r_y,\hat r_z)|\phi_\lambda\rangle&=\left ( \frac{\hbar^2}{2m\hat r^2}\left [l(l+1)+2\sum_{j=l}^{n-2}(\nu_j+1)-\alpha(\alpha+1)\right ]\right.\nonumber\\&+\left.\frac{e^2}{\lambda \hat r}(\alpha+1-\lambda)-\frac{e^2}{2\lambda^2 a_0}\right ) \hat r^{\alpha-l}P_h^l(\hat r_x,\hat r_y,\hat r_z)|\phi_\lambda\rangle.
\end{align}
Clearly, we must have again that $\alpha=\lambda-1$. To determine the $\nu_j$ terms, we must determine the factor $C$ when ${\mathcal H}(1)+\hbar^2\sum_{j=l}^{k-1}(\nu_j+1)/(m\hat r^2)$ moves to the right past $\hat B_r^\dagger(\nu_k)$, for each value of $k$ such that $l\le k\le n-2$. A simple calculation shows that
\begin{equation}
C_k=\nu_k(\nu_k+1)-2\sum_{j=l}^{k-1}(\nu_j+1).
\end{equation}
Starting from the first term, we must have $\nu_l=l$, otherwise $C_l\ne l(l+1)$. Given $\nu_l=l$, then we have $\nu_{l+1}$ must satisfy $\nu_{l+1}(\nu_{l+1}+1)-2(l+1)=l(l+1)$, or $\nu_{l+1}=l+1$ (because all $\nu_l$ are nonnegative). Continuing, in turn, we immediately see that we must have $\nu_j=j$ for the state to be an eigenstate. Then, we have
\begin{equation}
2\sum_{j=l}^{n-2}(j+1)=n(n-1)-l(l+1),
\end{equation}
which further tells us that we must have $\alpha=n-1$ again. Hence $\lambda=n$ and the energy eigenvalue is $-e^2/(2n^2a_0)$---the same value for every state in the chain that we can create, down to $l=0$. This completes the argument that we must choose consecutive integers in the chain of operators to yield an energy eigenstate.
The results are shown in Fig.~\ref{fig:energy} for the first three energy multiplets. We next establish these results rigorously, deriving all of the required details. A schematic of how the calculation is structured is given in Fig.~\ref{fig:schematic}.

We next establish the following identity, which involves multiple applications of the intertwining identity to move the Hamiltonian operator through the sequence of raising operators involving the product with $\lambda$ chosen from the $n-l-1$ factors $\{l,l+1,l+2,\cdots,n-3,n-2\}$ in order from left to right:
\begin{align}
&\hat{\mathcal H}(1)\left [ \hat B_r^\dagger(l)\hat B_r^\dagger(l+1)\cdots \hat B_r^\dagger(n-2)\right ]\nonumber\\
&~~~~~=\hat B_r^\dagger(l)\cdots \hat B_r^\dagger(n-2)\left [\hat {\mathcal H}(1)+\frac{\hbar^2}{m\hat r^2}\sum_{j=1}^{n-l-1}(n-j)\right ]\nonumber\\
&~~~~~+~\Biggr \{\left [ \hat B_r^\dagger(l)-\frac{2i\hbar}{\sqrt{2m}\hat r}\right ]\cdots \left [ \hat B_r^\dagger(n-2)-\frac{2i\hbar}{\sqrt{2m}\hat r}\right ]-\left [\hat B_r^\dagger(l)\cdots\hat B_r^\dagger(n-2)\right ]\Biggr \} \nonumber\\
&~~~~~~~\times\left [\hat T_\perp-\frac{\hbar^2l(l+1)}{2m\hat r^2}\right ]
\label{eq: main_intertwine}
\end{align}
We proceed by induction.
One can immediately verify that if we set $\lambda=l=n-2$, then Eq.~(\ref{eq: br_intertwine2}) is the base case for the above intertwining identity. We next assume it holds for the above case and proceed to show it also holds when we add the $\hat B_r^\dagger(l-1)$ operator to the product on the left. This yields
\begin{align}
\hat{\mathcal H}(1)\left [ \hat B_r^\dagger(l-1)\hat B_r^\dagger(l)\cdots \hat B_r^\dagger(n-2)\right ]&=\hat B_r^\dagger(l-1)\left [ \hat {\mathcal H}(1)+\frac{\hbar^2l}{m\hat r^2}\right ]\hat B_r^\dagger(l)\cdots \hat B_r^\dagger(n-2)\nonumber\\
&-\frac{2i\hbar}{\sqrt{2m}\hat r}\left (\hat T_\perp-\frac{\hbar^2(l-1)l}{2m\hat r^2}\right )\hat B_r^\dagger(l)\cdots\hat B_r^\dagger(n-2)\nonumber\\
\end{align}
after using Eq.~(\ref{eq: br_intertwine2}). Now we employ Eq.~(\ref{eq: main_intertwine}) along with (\ref{eq: br_tperp_interchange}) and (\ref{eq: r_tperp_interchange}) to find 
\begin{align}
\hat{\mathcal H}(1)\left [ \hat B_r^\dagger(l-1)\right .&\cdots \left .\hat B_r^\dagger(n-2)\right ]=\hat B_r^\dagger(l-1)\cdots\hat B_r^\dagger(n-2)\left [\hat {\mathcal H}(1)+\frac{\hbar^2}{m\hat r^2}\sum_{j=1}^{n-l-1}(n-j)\right ]\nonumber\\
&+\hat B_r^\dagger(l-1)\Biggr \{\left [ \hat B_r^\dagger(l)-\frac{2i\hbar}{\sqrt{2m}\hat r}\right ]\cdots \left [ \hat B_r^\dagger(n-2)-\frac{2i\hbar}{\sqrt{2m}\hat r}\right ]\nonumber\\
&-\left [\hat B_r^\dagger(l)\cdots\hat B_r^\dagger(n-2)\right ]\Biggr \} \left [\hat T_\perp-\frac{\hbar^2l(l+1)}{2m\hat r^2}\right ]\\
&+\hat B_r^\dagger(l-1)\left [ \hat B_r^\dagger(l)-\frac{2i\hbar}{\sqrt{2m}\hat r}\right ]\cdots \left [ \hat B_r^\dagger(n-2)-\frac{2i\hbar}{\sqrt{2m}\hat r}\right ]\frac{\hbar^2 l}{m\hat r^2}\nonumber\\
&-\frac{2i\hbar}{\sqrt{2m}\hat r}\left [ \hat B_r^\dagger(l)-\frac{2i\hbar}{\sqrt{2m}\hat r}\right ]\cdots \left [ \hat B_r^\dagger(n-2)-\frac{2i\hbar}{\sqrt{2m}\hat r}\right ]\left [\hat T_\perp-\frac{\hbar^2(l-1)l}{2m\hat r^2}\right ].\nonumber
\end{align}
The terms in the last two rows can be combined with those in the upper rows to finally yield
\begin{align}
&\hat{\mathcal H}(1)\left [ \hat B_r^\dagger(l-1)\cdots \hat B_r^\dagger(n-2)\right ]=\hat B_r^\dagger(l-1)\cdots\hat B_r^\dagger(n-2)\left [\hat {\mathcal H}(1)+\frac{\hbar^2}{m\hat r^2}\sum_{j=1}^{n-l}(n-j)\right ]\nonumber\\
&~~~~+\Biggr \{\left [ \hat B_r^\dagger(l-1)-\frac{2i\hbar}{\sqrt{2m}\hat r}\right ]\cdots \left [ \hat B_r^\dagger(n-2)-\frac{2i\hbar}{\sqrt{2m}\hat r}\right ]-\left [\hat B_r^\dagger(l-1)\cdots\hat B_r^\dagger(n-2)\right ]\Biggr \} \nonumber\\
&~~~~~~~~~~~~~~~~\times\left [\hat T_\perp-\frac{\hbar^2(l-1)l}{2m\hat r^2}\right ].
\label{eq: induction_intertwine_final}
\end{align}
This completes the inductive proof.

When constructing an ansatz for the eigenfunctions, we will find that the fact that the $\hat B_r$ operators have integer parameters will require the power of $\hat r$ and the parameter $\lambda$ to both be integers. So we choose $\lambda=n$ and the integer degree of the harmonic polynomial $l$ to label the eigenstate via
\begin{equation}
    |\psi_{nl}\rangle=\hat B_r^\dagger(l)
    \hat B_r^\dagger(l+1)\cdots\hat B_r^\dagger(n-2)\hat r^{n-l-1}P_h^l(\hat r_x,\hat r_y,\hat r_z)|\phi_{\lambda{=}n}\rangle,
\end{equation}
with the restriction that $0\le l\le n-1$ because harmonic polynomials are not defined for negative $l$, and our wavefunction ansatz restricts $l$ to be less than $n$.
Now we verify that it is indeed an eigenfunction and we also determine the eigenvalue.

This is the homestretch, but it still takes a number of steps. We use the intertwining relation (recalling all of the extra terms vanish when acting on the state to the right) and find that
\begin{align}
&\hat {\mathcal H}(1)|\psi_{nl}\rangle=\hat{\mathcal H}(1)\hat B_r^\dagger(l)
    \cdots\hat B_r^\dagger(n-2)\hat r^{n-l-1}P_h^l(\hat r_x,\hat r_y,\hat r_z)|\phi_{\lambda{=}n}\rangle\label{eq: eig1}\\
    &~=B_r^\dagger(l)
    \cdots\hat B_r^\dagger(n-2)\left [\hat {\mathcal H}(1)+\frac{\hbar^2}{2m\hat r^2}\{n(n-1)-l(l+1)\}\right ]\hat r^{n-l-1}P_h^l(\hat r_x,\hat r_y,\hat r_z)|\phi_{\lambda{=}n}\rangle,\nonumber
\end{align}
which follows from Eq.~(\ref{eq: main_intertwine}) and because the sum satisfies
\begin{equation}
    \sum_{j=1}^{n-l-1}(n-j)=n(n-l-1)-\frac{1}{2}(n-l-1)(n-l)=\frac{1}{2}[n(n-1)-l(l+1)].
\end{equation}

Next, we recall that $\hat {\mathcal H}(1)=\hat T-e^2/\hat r$ and focus on evaluating the term from the total kinetic energy as follows:
\begin{align}
    \hat T \hat r^{n-l-1}P_h^l(\hat r_x,\hat r_y,\hat r_z)|\phi_{\lambda{=}n}\rangle=
    [\hat T,\hat r^{n-l-1}P_h^l(\hat r_x,\hat r_y,\hat r_z)]|\phi_{\lambda{=}n}\rangle+
    \hat r^{n-l-1}P_h^l(\hat r_x,\hat r_y,\hat r_z)\hat T |\phi_{\lambda{=}n}\rangle.\nonumber\\
    \label{eq: T_harm_poly}
\end{align}
For the last term in Eq.~(\ref{eq: T_harm_poly}), we have 
\begin{equation}
    \hat T|\phi_{\lambda{=}n}\rangle=\left [\hat {\mathcal H}(n)+\frac{e^2}{n\hat r}\right ]|\phi_{\lambda{=}n}\rangle=\left [E(n)+\frac{e^2}{n\hat r}\right ]|\phi_{\lambda{=}n}\rangle.
\end{equation}
The first term in Eq.~(\ref{eq: T_harm_poly}) is evaluated by directly computing the commutator, which uses a number of the operator relations we already derived above [including the commutator in Eq.~(\ref{eq: rad_mom_commutes}), the identity in Eq.~(\ref{eq: tperp_identity}), and the action of $\hat p_r$ on $|\phi_\lambda\rangle$ in Eq.~(\ref{eq: pr_state})]. The result becomes
\begin{align}
    &\left [\hat T,\hat r^{n-l-1}P_h^l(\hat r_x,\hat r_y,\hat r_z)\right ]|\phi_{\lambda{=}n}\rangle=\left [\hat T_\perp,\hat r^{n-l-1}P_h^l(\hat r_x,\hat r_y,\hat r_z)\right ]|\phi_{\lambda{=}n}\rangle\nonumber\\
    &~~~~~~~~~~~~~~~~~~~~~~~~~~~~~~~~~~~~~~~~~~~~~~~~~~~~~+\left [\frac{\hat p_r^2}{2m},\hat r^{n-l-1}P_h^l(\hat r_x,\hat r_y,\hat r_z)\right ]|\phi_{\lambda{=}n}\rangle\nonumber\\
    &~~~~~~~~~~~~~~~~~~~~~~~~=\frac{\hbar^2 l(l+1)}{2m\hat r^2}\hat r^{n-l-1}P_h^l(\hat r_x,\hat r_y,\hat r_z)|\phi_{\lambda{=}n}\rangle+\frac{P_h^l(\hat r_x,\hat r_y,\hat r_z)}{\hat r^l}\left [\frac{\hat p_r^2}{2m},\hat r^{n-1}\right ]|\phi_{\lambda{=}n}\rangle\nonumber\\
    &=\Biggr [ \frac{\hbar^2\{l(l+1)-(n-2)(n-1)\}}{2m\hat r^2}+\frac{\hbar^2}{m\hat r}\left ( \frac{1}{na_0}-\frac{1}{\hat r}\right )(n-1)\Biggr ] \hat r^{n-l-1}P_h^l(\hat r_x,\hat r_y,\hat r_z)|\phi_{\lambda{=}n}\rangle.
\end{align}
Note that it is at this point, where we require $\lambda=n$. If this was not the case, the term proportional to $\hbar^2/(m\hat r a_0)=e^2/\hat r$ in the last equality would have a coefficient of $(n-1)/\lambda$ instead of $1-1/n$. We only obtain an eigenstate if the coefficient is exactly $1-1/n$.
Combining all of these results then gives
\begin{align}
\hat T\hat r^{n-l-1}P_h^l(\hat r_x,\hat r_y,\hat r_z)|\phi_{\lambda{=}n}\rangle&=\Biggr [ \frac{\hbar^2\{l(l+1)-n(n-1)\}}{2m\hat r^2}+\frac{e^2}{\hat r}+E(n)\Biggr ]\nonumber\\
&~~~~~\times\hat r^{n-l-1}P_h^l(\hat r_x,\hat r_y,\hat r_z)|\phi_{\lambda{=}n}\rangle.
\end{align}
Introducing this result into Eq.~(\ref{eq: eig1}) finally shows that 
\begin{equation}
    \hat{\mathcal H}(1)|\psi_{nl}\rangle=E(n)|\psi_{nl}\rangle,
\end{equation}
which is the eigenvalue-eigenvector relation we wanted to establish.

We have not yet considered the normalization of the wavefunction. To do this, we must compute the norm
\begin{align}
\label{eq: norm}
    \langle \psi_{nl}|\psi_{nl}\rangle&=\langle \phi_{\lambda{=}n}|P_h^l(\hat r_x,\hat r_y,\hat r_z)\hat r^{n-l-1}\hat B_r(n-2)\cdots\hat B_r(l)\\
    &~~~\times\hat B_r^\dagger(l)\cdots \hat B_r^\dagger(n-2)\hat r^{n-l-1}P_h^l(\hat r_x,\hat r_y,\hat r_z)|\phi_{\lambda{=}n}\rangle.\nonumber
\end{align}
We replace $\hat B_r(l)\hat B_r^\dagger(l)=\hat{\mathcal H}(1)-\hat T_\perp-E(l+1)+\hbar^2(l+1)(l+2)/(2m\hat r^2)$ in the center and then commute the terms through. Using the results we have derived in computing the eigenfunction above, a careful calculation shows that this result is equal to the factor $[E(n)-E(l+1)]$ multiplied by the expression in Eq.~(\ref{eq: norm}) with the middle two factors $\hat B_r(l)\hat B_r^\dagger(l)$ removed. Continuing in this fashion, we find that
\begin{equation}
    \langle \psi_{nl}|\psi_{nl}\rangle=
    \prod_{j=l+1}^{n-1}\left [E(n)-E(j)\right ]\langle \phi_{\lambda{=}n}|\hat r^{2n-2l-2}\left [P_h^l(\hat r_x,\hat r_y,\hat r_z)\right ]^2|\phi_{\lambda{=}n}\rangle.
\end{equation}
We assume that the initial state $\hat r^{n-1}|\phi_n\rangle$ is normalized and we choose the harmonic polynomials such that
\begin{equation}
    \langle \phi_{\lambda{=}n}|\hat r^{2n-2l-2}\left [P_h^l(\hat r_x,\hat r_y,\hat r_z)\right ]^2|\phi_{\lambda{=}n}\rangle=1.
\end{equation}
Since $P_h^l(\hat r_x,\hat r_y,\hat r_z)/\hat r^l$ is a linear combination of the conventional spherical harmonics, the above relation simply says the angular part of the wavefunction is normalized (given that the radial part was already normalized).
Then we find that the normalized eigenstates
are
\begin{equation}
\label{eq: wf_norm}
    |\psi_{nl}\rangle_{\rm norm}=\frac{(\sqrt{2a_0})^{n-l-1}}{e^{n-l-1}\sqrt{\prod_{j=l+1}^{n-1}\left ( \frac{1}{j^2}-\frac{1}{n^2}\right )}}
    \hat B_r^\dagger(l)\cdots \hat B_r^\dagger(n-2)\hat r^{n-l-1}P_h^l(\hat r_x,\hat r_y,\hat r_z)|\phi_{\lambda{=}n}\rangle,
\end{equation}
with the normalization factor for the auxiliary Hamiltonian ground state satisfying
\begin{equation}
    \langle \hat r_x{=}0,\hat r_y{=}0,\hat r_z{=}0| \phi_{\lambda{=}n}\rangle=\left (\frac{2}{na_0}\right )^{n+\frac{1}{2}}\frac{1}{\sqrt{(2n)!}}.
    \label{eq: norm_coord}
\end{equation}
The last result requires calculus to compute, and is the standard normalization condition. Note that the $e$ appearing above is the electric charge of the proton.

This derivation has produced all of the bound states of Hydrogen labeled by the principle quantum number $n$ and the total angular momentum $l$. But, because we are working with harmonic polynomials, we are not generically in an eigenstate of the $z$-component of angular momentum. We have the restriction that $0\le l<n$, which implies that there is a degeneracy across $l$ and we have determined the well-known bound-state energy $E(n)=-e^2/(2n^2a_0)$. All of this is as expected, but it used no angular momentum and no calculus (except for the final normalization step)!

\section{The wavefunction in coordinate space}

Now we focus on determining the coordinate representation for the unnormalized wavefunction given by the overlap
\begin{equation}
    \psi_{nl}(r_x,r_y,r_z)=\langle r_x,r_y,r_z|\psi_{nl}\rangle.
\end{equation}
This wavefunction can also be found algebraically, employing no calculus, as we show next.

To begin, we first determine $\langle r_x,r_y,r_z|\psi_{nn-1}\rangle$, which satisfies
\begin{equation}
    \langle r_x,r_y,r_z|\psi_{nn-1}\rangle=\langle r_x,r_y,r_z|P_h^{n-1}(\hat r_x,\hat r_y,\hat r_z)|\phi_{\lambda{=}n}\rangle.
\end{equation}
We can immediately act the harmonic polynomial on the coordinate state to the left to obtain
\begin{equation}
    \langle r_x,r_y,r_z|\psi_{nn-1}\rangle=P_h^{n-1}(r_x,r_y,r_z)\langle r_x,r_y,r_z|\phi_{\lambda{=}n}\rangle.
\end{equation}
Then we use the techniques from Eqs.~(\ref{eq: translation}), (\ref{eq: gs_wf}) and the Appendix to find the result generalizing Eq.~(\ref{eq: gs_final}) as follows:
\begin{equation}
    \langle r_x,r_y,r_z|\psi_{nn-1}\rangle=P_h^{n-1}(r_x,r_y,r_z)e^{-\frac{r}{na_0}}\langle r_x{=}0,r_y{=}0,r_z{=}0|\phi_{\lambda{=}n}\rangle,
\end{equation}
where $\langle r_x{=}0,r_y{=}0,r_z{=}0|\phi_{\lambda{=}n}\rangle$ is the normalization constant in Eq.~(\ref{eq: norm_coord}). This requires calculus to determine in the usual fashion. 
Note that this result might not be in a form that is easily recognized, since we do not have a conventional spherical harmonic factor. Rest assured, this result is the same as the conventional result, with a radial wavefunction behaving like $r^l$ near the origin. We simply rewrite it as 
\begin{equation}
    \langle \vec{r}|\psi_{nn-1}\rangle_{\rm norm}=\left (\frac{2}{na_0}\right )^{n+\frac{1}{2}}\frac{1}{\sqrt{(2n)!}}r^{n-1}e^{-\frac{r}{na_0}}\frac{P_h^{n-1}(r_x,r_y,r_z)}{r^{n-1}},
    \label{eq: r_aux_wf}
\end{equation}
where the last term is a properly normalized linear combination of the conventional spherical harmonics $Y_{lm}(\theta,\phi)$; that normalization is with respect to the angular coordinates.

For the general wavefunction $|\psi_{nl}\rangle$ we must work with the $\hat B_r(\lambda)$ operators. The general wavefunction satisfies
\begin{equation}
\langle\vec{r}|\psi_{nl}\rangle=\langle\vec{r}|\hat B_r^\dagger(l)\cdots\hat B_r^\dagger(n-2)\hat r^{n-l-1}P_h^l(\hat r_x,\hat r_y,\hat r_z)|\phi_{\lambda{=}n}\rangle.
\end{equation}
The first step is to note that we can move the ``spherical harmonic'' factor $P_h^l(\hat r_x,\hat r_y, \hat r_z)/\hat r^l$ all the way to the left and have it act on the position eigenstate because that term commutes with all $\hat{B}_r^\dagger$ operators. Doing so gives
\begin{equation}
\langle\vec{r}|\psi_{nl}\rangle=\frac{P_h^l(r_x,r_y,r_z)}{r^l}\langle\vec{r}|\hat B_r^\dagger(l)\cdots\hat B_r^\dagger(n-2)\hat r^{n-1}|\phi_{\lambda{=}n}\rangle,
\label{eq: rwf2}
\end{equation}
and we focus all of our efforts on evaluating the remaining matrix element. Note that evaluating a generic term, like $\hat B_r^\dagger(\lambda)\hat r^\mu|\phi_n\rangle$ can always be carried out by using Eq.~(\ref{eq: pr_state}) and the relation
\begin{align}
\hat B_r^\dagger(\lambda)\hat r^\mu|\phi_{\lambda{=}n}\rangle&=
\left [ \hat B_r^\dagger(\lambda),\hat r^\mu\right ]|\phi_n\rangle
+\hat r^\mu\hat B_r^\dagger(\lambda)|\phi_n\rangle\nonumber\\
&=\frac{i\hbar}{\sqrt{2m}}\left [-\mu\hat r^{\mu-1}+\frac{\hat r^\mu}{na_0}-\hat r^{\mu-1}+\frac{\hat r^\mu}{(\lambda+1)a_0}-(\lambda+1)\hat r^{\mu-1}\right ]|\phi_{\lambda{=}n}\rangle.
\end{align}
So each time a $\hat B_r^\dagger$ operator acts on a power of $\hat r$ multiplied by an auxiliary Hamiltonian ground state, it produces a number times the same power plus another number times the power minus one. Hence, the remaining matrix element in Eq.~(\ref{eq: rwf2}) is a polynomial in $r$ of degree $n-1$ (that has a lowest power of $n-l-1$) and is multiplied by the auxiliary Hamiltonian ground-state wavefunction (exponential function) in Eq.~(\ref{eq: r_aux_wf}).

We call this (unnormalized) radial wavefunction $R_{nl}(r)$, which satisfies
\begin{equation}
        R_{nl}(r)=(-i)^{n-l-1}\langle \vec{r}|\hat B_r^\dagger(l)\cdots\hat B_r^\dagger(n-2)
    \hat r^{n-1}|\phi_{\lambda{=}n}\rangle.
    \label{eq: rnl_def}
\end{equation}
The factor $(i)^{n-l-1}$ is just a phase factor, which is introduced so that this function agrees with the conventional definition of the radial wavefunction.
We can immediately compute the first few of these functions. Namely, we find that the first function in each series satisfies
\begin{equation}
    R_{nn-1}=\left (\frac{2}{na_0}\right )^{n+\frac{1}{2}}\frac{1}{\sqrt{(2n)!}}r^{n-1}e^{-\frac{r}{na_0}},
\end{equation}
the second satisfies
\begin{equation}
    R_{nn-2}(r)=\left (\frac{2}{na_0}\right )^{n+\frac{1}{2}}\frac{1}{\sqrt{(2n)!}}\frac{\hbar(2n-1)}{\sqrt{2m}}\left [
    \frac{1}{(n-1)(na_0)}r^{n-1}-r^{n-2} \right ]e^{-\frac{r}{na_0}}
    \label{eq: second}
\end{equation}
and the third
\begin{align}
R_{nn-3}(r)&=\left (\frac{2}{na_0}\right )^{n+\frac{1}{2}}\frac{1}{\sqrt{(2n)!}}\frac{\hbar^2(2n-1)}{2m}\\
&\times\left [
    \frac{(2n-2)}{(n-1)(n-2)(na_0)^2}r^{n-1}\right .-\left .
    \frac{(2n-2)(2n-3)}{(n-1)(n-2)(na_0)}r^{n-2}+(2n-3)r^{n-3} \right ]e^{-\frac{r}{na_0}}.\nonumber
\end{align}

The strategy we employ to determine the polynomial in the general case is to employ Rodrigues formulas, but here expressed in an operator language in terms of the radial momentum operator, instead of derivatives. The proof of this result requires a proof by induction. It is key to note that the identity that is derived is not a pure operator identity. It only holds for the string of operators acting on the specific energy eigenstate. This will become clear as we work it out. The wavefunction can also be found by explicitly computing the polynomials that are created from the string of $\hat{B}_r^\dagger$ operators. This calculation was completed elsewhere~\cite{rushka}.

To start, we need to be sure we have the proper Rodrigues formula for the Laguerre polynomials. 
We define the order $k$ polynomial via
\begin{equation}
    L_{k}^{\alpha}(x)=\sum_{j=0}^{k}\frac{(-1)^j}{j!}\begin{pmatrix}k+\alpha\\k-j\end{pmatrix}x^j,
\end{equation}
where $k$ is a non-negative integer. This is the modern convention for the Laguerre polynomials (and used by many texts, such as the one by Powell and Crasemann~\cite{powell_crasemann}); it is different from the one used by Schiff~\cite{schiff}, which is the more common notation employed in quantum mechanics textbooks. The Rodrigues formula is
\begin{equation}
    L_k^\alpha(x)=\frac{x^{-\alpha}}{k!}e^x\frac{d^k}{dx^k}\left (x^{k+\alpha}e^{-x}\right ).
\end{equation}
One can directly check these two definitions are exactly the same.

Before we can begin with our derivation, we need to evaluate some operator identities. We start from the Hadamard lemma 
\begin{equation}
    e^{\hat{A}}\hat{B}e^{-\hat{A}}=\hat{B}+[\hat{A},\hat{B}]+\frac{1}{2}[\hat{A},[\hat{A},\hat{B}]]+\cdots
\end{equation}
where the summation includes a sequence of multiple nested commutators, with the $n$-fold nested commutator weighted by $\frac{1}{n!}$. Then, one immediately finds that
\begin{equation}
    e^{a\hat{r}}\hat{p}_re^{-a\hat{r}}=\hat{p_r}+i\hbar a
\end{equation}
and
\begin{equation}
    \hat{r}^b\hat{p}_r\hat{r}^{-b}=e^{b\ln\hat{r}}\hat{p}_r e^{-b\ln\hat{r}}=\hat{p}_r+i\hbar \frac{b}{\hat{r}}.
\end{equation}
The second identity can also be derived from $[\hat{r}^b,\hat{p}_r]=i\hbar b \hat{r}^{b-1}$, by multiplying from the right by $\hat{r}^{-b}$.

Now, we are ready to work out the base case for the proof by induction. We need to convert 
\begin{equation}
    \hat{B}_r^\dagger(n-2)\hat{r}^{n-1}|\phi_{\lambda{=}n}\rangle=\frac{1}{\sqrt{2m}}\left[\hat{p}_r+\frac{i\hbar}{(n-1)a_0}-\frac{i\hbar(n-1)}{\hat{r}}\right ]\hat{r}^{n-1}|\phi_{\lambda{=}n}\rangle
\end{equation}
into a Rodrigues formula for an operator acting on $|\phi_n\rangle$.
Using the two operator identities above, we can rewrite this as
\begin{equation}
    \hat{B}_r^\dagger(n-2)\hat{r}^{n-1}|\phi_{\lambda{=}n}\rangle=\frac{1}{\sqrt{2m}}\hat{r}^{-(n-1)}e^{\frac{\hat{r}}{(n-1)a_0}}\hat{p}_re^{-\frac{\hat{r}}{(n-1)a_0}}\hat{r}^{2(n-1)}|\phi_{\lambda{=}n}\rangle.
\end{equation}
The power of $\hat{r}$ on the right-hand side, just before the state is $2(n-1)$.
We next ``multiply by one'' in two places and then recognize and evaluate a Hadamard lemma:
\begin{align}
    \hat{B}_r^\dagger(n-2)\hat{r}^{n-1}|\phi_{\lambda{=}n}\rangle&=\frac{1}{\sqrt{2m}}\hat{r}^{-(n-1)}\underbrace{e^{\frac{\hat{r}}{na_0}}e^{-\frac{\hat{r}}{na_0}}}_{=1}e^{\frac{\hat{r}}{(n-1)a_0}}\hat{p}_r e^{-\frac{\hat{r}}{(n-1)a_0}}\underbrace{e^{\frac{\hat{r}}{na_0}}e^{-\frac{\hat{r}}{na_0}}}_{=1}\hat{r}^{2(n-1)}|\phi_{\lambda{=}n}\rangle\nonumber\\
    &=\frac{1}{\sqrt{2m}}\hat{r}^{-(n-1)}e^{\frac{\hat{r}}{na_0}}\underbrace{e^{-\frac{\hat{r}}{a_0}\left (\frac{1}{n}-\frac{1}{n-1}\right )}\hat{p}_r e^{\frac{\hat{r}}{a_0}\left (\frac{1}{n}-\frac{1}{n-1}\right )}}_{\text{Hadamard lemma}}e^{-\frac{\hat{r}}{na_0}}\hat{r}^{2(n-1)}|\phi_{\lambda{=}n}\rangle\nonumber\\
    &=\frac{1}{\sqrt{2m}}\hat{r}^{-(n-1)}e^{\frac{\hat{r}}{na_0}}\hat{p}_re^{-\frac{\hat{r}}{na_0}}\hat{r}^{2(n-1)}|\phi_n\rangle+\frac{1}{\sqrt{2m}}\frac{i\hbar}{n(n-1)a_0}\hat{r}^{n-1}|\phi_{\lambda{=}n}\rangle.
\end{align}
We construct another Hadamard lemma with powers of $\hat{r}$:
\begin{align}
    \hat{B}_r^\dagger(n-2)|\phi_{\lambda{=}n}\rangle&=\frac{1}{\sqrt{2m}}\hat{r}^{-(n-2)}e^{\frac{\hat{r}}{na_0}}\underbrace{\frac{1}{\hat{r}}\hat{p}_r\hat{r}}_{\text{Hadamard}}e^{-\frac{\hat{r}}{na_0}}\hat{r}^{2n-3}|\phi_{\lambda{=}n}\rangle\nonumber\\
    &~~~~~+\frac{1}{\sqrt{2m}}\frac{i\hbar}{n(n-1)a_0}\hat{r}^{n-1}|\phi_{\lambda{=}n}\rangle\nonumber\\
    &=\frac{1}{\sqrt{2m}}\hat{r}^{-(n-2)}e^{\frac{\hat{r}}{na_0}}\hat{p}_re^{-\frac{\hat{r}}{na_0}}\hat{r}^{2n-3}|\phi_{\lambda{=}n}\rangle\nonumber\\
    &~~~~~+\frac{1}{\sqrt{2m}}\frac{1}{2(n-1)}\left (\frac{2i\hbar}{na_0}-\frac{2i\hbar(n-1)}{\hat{r}}\right )\hat{r}^{n-1}|\phi_{\lambda{=}n}\rangle
\end{align}
Next, using Eq.~(\ref{eq: pr_state}), we replace $i\hbar\left (\frac{1}{na_0}-\frac{n}{\hat{r}}\right)$ acting on $\hat{r}^{n-1}|\phi_{\lambda{=}n}\rangle$ by $\hat{p}_r$ acting on the same state. Hence, we have
\begin{align}
    \hat{B}_r^\dagger(n-2)|\phi_{\lambda{=}n}\rangle&=\frac{1}{\sqrt{2m}}\hat{r}^{-(n-2)}e^{\frac{\hat{r}}{na_0}}\hat{p}_re^{-\frac{\hat{r}}{na_0}}\hat{r}^{2n-3}|\phi_{\lambda{=}n}\rangle\nonumber\\
    &~~~~~+\frac{1}{\sqrt{2m}}\frac{1}{2(n-1)}\left (\hat{p}_r+\frac{i\hbar}{na_0}-\frac{i\hbar(n-2)}{\hat{r}}\right )\hat{r}^{n-1}|\phi_{\lambda{=}n}\rangle\nonumber\\
    &=\left (1+\frac{1}{2(n-1)}\right )\frac{1}{\sqrt{2m}}\hat{r}^{-(n-2)}e^{\frac{\hat{r}}{na_0}}\hat{p}_re^{-\frac{\hat{r}}{na_0}}\hat{r}^{2n-3}|\phi_{\lambda{=}n}\rangle
\end{align}
after we use the Hadamard identities again on the second term. Note how we must use properties of the state to complete this derivation. This is why it is not a pure operator identity, but instead is an identity that requires the chain of operators to act on the state, in order for it to hold.

So, we have shown that
\begin{equation}
    \hat{B}_r^\dagger(n-2)\hat{r}^{n-1}|\phi_{\lambda{=}n}\rangle=\frac{2n-1}{2n-2}\frac{1}{\sqrt{2m}}\hat{r}^{-(n-2)}e^{\frac{\hat{r}}{na_0}}\hat{p}_re^{-\frac{\hat{r}}{na_0}}\hat{r}^{2n-3}|\phi_{\lambda{=}n}\rangle.
\end{equation}
This is our base case for the proof by induction. Determining the general case requires working out some additional examples to determine the general pattern for how this works. We find the induction hypothesis is
\begin{align}
    &\hat{B}_r^\dagger(l)\hat{B}_r^\dagger(l+1)\cdots\hat{B}_r(n-3)\hat{B}_r^\dagger(n-2)\hat{r}^{n-1}|\phi_{\lambda{=}n}\rangle\nonumber\\
    &~~~~~=\frac{(2n-1)!\,l!}{2^{n-l-1}(n+l)!\,(n-1)!}\frac{1}{\sqrt{(2m)^{n-l-1}}}\,\hat{r}^{-l}e^{\frac{\hat{r}}{na_0}}(\hat{p}_r)^{n-l-1}e^{-\frac{\hat{r}}{na_0}}\hat{r}^{n+l-1}|\phi_{\lambda{=}n}\rangle.
    \label{eq: laguerre-identity}
\end{align}
In order to make the equations a bit less cumbersome, we define
\begin{equation}
    c(l)=\frac{(2n-1)!\,l!}{2^{n-l-1}(n+l)!\,(n-1)!}\frac{1}{\sqrt{(2m)^{n-l-1}}}.
\end{equation}
Having established the base case already for angular momentum equal to $n-2$, we assume it holds for all angular momentum down to $l$ and use this information to prove the result for angular momentum $l-1$. So, our starting point is
\begin{align}
    &\hat{B}_r^\dagger(l-1)\hat{B}_r^\dagger(l)\cdots\hat{B}_r^\dagger(n-2)\hat{r}^{n-1}|\phi_{\lambda{=}n}\rangle\nonumber\\
    &~~~~~=\frac{1}{\sqrt{2m}}\left (\hat{p}_r+\frac{i\hbar}{la_0}-\frac{i\hbar l}{\hat{r}}\right )c(l)\hat{r}^{-l}e^{\frac{\hat{r}}{na_0}}(\hat{p}_r)^{n-l-1}e^{-\frac{\hat{r}}{na_0}}\hat{r}^{n+l-1}|\phi_{\lambda{=}n}\rangle\nonumber\\
    &~~~~~=\frac{c(l)}{\sqrt{2m}}\left (\hat{r}^{-l}e^{\frac{\hat{r}}{la_0}}\hat{p}_r e^{-\frac{\hat{r}}{la_0}}\hat{r}^l\right )\hat{r}^{-l}e^{\frac{\hat{r}}{na_0}}(\hat{p}_r)^{n-l-1}e^{-\frac{\hat{r}}{na_0}}\hat{r}^{n+l-1}|\phi_{\lambda{=}n}\rangle
\end{align}
The $\hat{r}^l$ term cancels with the $\hat{r}^{-l}$ term in between the radial momentum operators. We combine the two exponential operators and move the single radial momentum operator through the exponential term to join with the remaining radial momentum operator. This is done using our multiply by one and Hadamard lemma identities yielding
\begin{align}
    &\hat{B}_r^\dagger(l-1)\hat{B}_r^\dagger(l)\cdots\hat{B}_r^\dagger(n-2)\hat{r}^{n-1}|\phi_{\lambda{=}n}\rangle\nonumber\\
    &~~~~~=\frac{c(l)}{\sqrt{2m}}\hat{r}^{-l}e^{\frac{\hat{r}}{na_0}}(\hat{p}_r)^{n-l}e^{-\frac{\hat{r}}{na_0}}\hat{r}^{n+l-1}|\phi_{\lambda{=}n}\rangle\nonumber\\
    &~~~~~+\frac{c(l)}{\sqrt{2m}}\hat{r}^{-l}e^{\frac{\hat{r}}{na_0}}(\hat{p}_r)^{n-l-1}e^{-\frac{\hat{r}}{na_0}}\hat{r}^{n+l-1}\frac{i\hbar}{a_0}\frac{n-l}{nl}|\phi_{\lambda{=}n}\rangle.
\end{align}
Next, we take one factor of $\hat{r}$ from the right, and commute it through the radial momentum terms in the first term on the right-hand side. This gives
\begin{align}
    &\hat{B}_r^\dagger(l-1)\hat{B}_r^\dagger(l)\cdots\hat{B}_r^\dagger(n-2)\hat{r}^{n-1}|\phi_{\lambda{=}n}\rangle\nonumber\\
    &~~~~~=\frac{c(l)}{\sqrt{2m}}\hat{r}^{-l+1}e^{\frac{\hat{r}}{na_0}}(\hat{p}_r)^{n-l}e^{-\frac{\hat{r}}{na_0}}\hat{r}^{n+l-2}|\phi_{\lambda{=}n}\rangle\nonumber\\
    &~~~~~+\frac{c(l)}{\sqrt{2m}}\hat{r}^{-l}e^{\frac{\hat{r}}{na_0}}(\hat{p}_r)^{n-l-1}e^{-\frac{\hat{r}}{na_0}}\hat{r}^{n+l-1}\left (\frac{i\hbar}{a_0}\frac{n-l}{nl}-(n-l)\frac{i\hbar}{\hat{r}}\right )|\phi_{\lambda{=}n}\rangle.\nonumber\\
    &~~~~~=\frac{c(l)}{\sqrt{2m}}\hat{r}^{-l+1}e^{\frac{\hat{r}}{na_0}}(\hat{p}_r)^{n-l}e^{-\frac{\hat{r}}{na_0}}\hat{r}^{n+l-2}|\phi_{\lambda{=}n}\rangle\nonumber\\
    &~~~~~+\frac{n-l}{2l}\frac{c(l)}{\sqrt{2m}}\hat{r}^{-l}e^{\frac{\hat{r}}{na_0}}(\hat{p}_r)^{n-l-1}e^{-\frac{\hat{r}}{na_0}}\hat{r}^{n+l-1}\left (\hat{p}_r+\frac{i\hbar}{na_0}-(2l-1)\frac{i\hbar}{\hat{r}}\right )|\phi_{\lambda{=}n}\rangle,
\end{align}
where the last line arises by factoring out $\frac{n-l}{2l}$ and replacing $\left (\frac{i\hbar}{na_0}-\frac{i\hbar}{\hat{r}}\right)|\phi_{\lambda{=}n}\rangle=\hat{p}_r|\phi_{\lambda{=}n}\rangle$ in the last term. Next, we move the factor in the parenthesis though the exponential and power factors using the Hadamard lemma. The radial momentum term can then join the other radial momentum terms. This gives 
\begin{align}
    &\hat{B}_r^\dagger(l-1)\hat{B}_r^\dagger(l)\cdots\hat{B}_r^\dagger(n-2)\hat{r}^{n-1}|\phi_{\lambda{=}n}\rangle\nonumber\\
    &~~~~~=\frac{c(l)}{\sqrt{2m}}\hat{r}^{-l+1}e^{\frac{\hat{r}}{na_0}}(\hat{p}_r)^{n-l}e^{-\frac{\hat{r}}{na_0}}\hat{r}^{n+l-2}|\phi_{\lambda{=}n}\rangle\nonumber\\
    &~~~~~+\frac{n-l}{2l}\frac{c(l)}{\sqrt{2m}}\hat{r}^{-l}e^{\frac{\hat{r}}{na_0}}(\hat{p}_r)^{n-l}e^{-\frac{\hat{r}}{na_0}}\hat{r}^{n+l-1}|\phi_{\lambda{=}n}\rangle\nonumber\\
    &~~~~~+\frac{n-l}{2l}\frac{c(l)}{\sqrt{2m}}\hat{r}^{-l}e^{\frac{\hat{r}}{na_0}}(\hat{p}_r)^{n-l-1}\left (i\hbar\frac{n-l}{\hat{r}}\right )e^{-\frac{\hat{r}}{na_0}}\hat{r}^{n+l-1}|\phi_{\lambda{=}n}\rangle.
\end{align}
The final step is to take one factor of $\hat{r}$ from the right and commute it through the radial momentum terms in the middle term on the right-hand side. The commutator cancels the third term, and the second term becomes identical to the first, except for the constant factor out front. Adding those two terms gives us the final result
\begin{align}
    &\hat{B}_r^\dagger(l-1)\hat{B}_r^\dagger(l)\cdots\hat{B}_r^\dagger(n-2)\hat{r}^{n-1}|\phi_{\lambda{=}n}\rangle\nonumber\\
    &~~~~~=\frac{c(l)}{\sqrt{2m}}\frac{n+l}{2l}\hat{r}^{-l+1}e^{\frac{\hat{r}}{na_0}}(\hat{p}_r)^{n-l}e^{-\frac{\hat{r}}{na_0}}\hat{r}^{n+l-2}|\phi_{\lambda{=}n}\rangle.
\end{align}
The coefficient in front is $c(l-1)$ because $c(l-1)=\frac{c(l)}{\sqrt{2m}}\frac{n+l}{2l}$. This completes the proof. Hence, Eq.~(\ref{eq: laguerre-identity}) holds.

We still need to show that this formula is the Rodrigues formula for the Laguerre polynomial. Key to this derivation is constructing a state $|\psi_n\rangle$ from $|\phi_{\lambda{=}n}\rangle$ that is annihilated by $\hat{p}_r$. Using Eq.~(\ref{eq: pr_state}), we see that $|\psi_n\rangle=e^{\frac{\hat{r}}{na_0}}\frac{1}{\hat{r}}|\phi_{\lambda{=}n}\rangle$ because
\begin{align}
    \hat{p}_r|\psi_n\rangle&=\hat{p_r}e^{\frac{\hat{r}}{na_0}}\frac{1}{\hat{r}}|\phi_{\lambda{=}n}\rangle\nonumber\\
    &=\left [\hat{p}_r,e^{\frac{\hat{r}}{na_0}}\frac{1}{\hat{r}}\right ]|\phi_{\lambda{=}n}\rangle+e^{\frac{\hat{r}}{na_0}}\frac{1}{\hat{r}}\hat{p}_r|\phi_{\lambda{=}n}\rangle\nonumber\\
    &=e^{\frac{\hat{r}}{na_0}}\frac{1}{\hat{r}}\left (\underbrace{-\frac{i\hbar}{na_0}+\frac{i\hbar}{\hat{r}}}_{\text{commutator}}+\underbrace{\frac{i\hbar}{na_0}-\frac{i\hbar}{\hat{r}}}_{\text{state}}\right )|\phi_{\lambda{=}n}\rangle=0.
\end{align}
We use a multiply by one to introduce this state (via $|\phi_{\lambda{=}n}\rangle=e^{-\frac{\hat{r}}{na_0}}\hat{r}|\psi_n\rangle$), then the powers of momentum operators can be replaced by nested commutators, because
\begin{equation}
    \hat{p}_rf(\hat{r})|\psi_n\rangle=[\hat{p}_r,f(\hat{r})]|\psi_n\rangle,
\end{equation}
since $\hat{p}_r|\psi_n\rangle=0$. This allows us to reduce the operator Rodrigues formula in Eq.~(\ref{eq: laguerre-identity}) to a series of nested commutators via
\begin{align}
    &\hat{B}_r^\dagger(l)\cdots\hat{B}_r^\dagger(n-2)\hat{r}^{n-1}|\phi_{\lambda{=}n}\rangle\nonumber\\
    &~~~~~=c(l)\hat{r}^{-l}e^{\frac{\hat{r}}{na_0}}(\hat{p}_r)^{n-l-1}e^{-\frac{2\hat{r}}{na_0}}\hat{r}^{n+l}|\psi_n\rangle\\
    &~~~~~=c(l)\hat{r}^{l+1}e^{-\frac{\hat{r}}{na_0}}\left ( \hat{r}^{-(2l+1)}e^{\frac{2\hat{r}}{na_0}}\left [\hat{p}_r,\left [\hat{p}_r\left [,\cdots,\left [\hat{p}_r,e^{-\frac{2\hat{r}}{na_0}}\hat{r}^{n+l}\right]\cdots\right]\right]\right ]_{n-l-1}\right )|\psi_n\rangle.\nonumber
\end{align}
Since the commutator of $\hat{p}_r$ with $f(\hat{r})$ acts like $-i\hbar$ times a derivative with respect to $\hat{r}$ of $f$, this is nearly in the form of a Rodrigues formula. We only need to ensure that the powers of $\hat{r}$ are multiplied by the right constant to make them dimensionless and equal to the argument of the Laguerre polynomial (which is determined by the argument of the exponential function); we also need to ensure the derivative term is likewise dimensionless. Hence, the operator form of the Rodrigues formula for the associated Laguerre polynomial becomes
\begin{align}
    &\left (\frac{ina_0}{2\hbar}\right )^{n-l-1}\left(\frac{2\hat{r}}{na_0}\right)^{-(2l+1)}e^{\frac{2\hat{r}}{na_0}}\left [\hat{p}_r,\left [\hat{p}_r\left [,\cdots,\left [\hat{p}_r,e^{-\frac{2\hat{r}}{na_0}}\left (\frac{2\hat{r}}{na_0}\right )^{n+l}\right]\cdots\right]\right]\right ]_{n-l-1}\nonumber\\
    &~~~~~=(n-l-1)!\,L_{n-l-1}^{2l+1}\left (\frac{2\hat{r}}{na_0}\right).
\end{align}
Defining 
\begin{equation}
    \hat{\rho}=\frac{2\hat{r}}{na_0},
\end{equation}
our final operator-state identity is
\begin{align}
    &\hat{B}_r^\dagger(l)\cdots\hat{B}_r^\dagger(n-2)\hat{r}^{n-1}|\phi_{\lambda{=}n}\rangle\nonumber\\
    &~~~~~=\frac{(2n-1)!\,l!\,(n-l-1)!}{2^{n-l-1}(n+l)!\,(n-1)!}\left (\frac{\hbar}{i\sqrt{2m}}\right )^{n-l-1}\left(\frac{na_0}{2}\right)^{l}L_{n-l-1}^{2l+1}(\hat{\rho})\hat{\rho}^{l}|\phi_{\lambda{=}n}\rangle.
\end{align}

To finish the calculation of the wavefunction in real space, we multiply the operator expression from the right by the position-space bra $\langle r_x,r_y,r_z|$, which allows us to replace $\hat{\rho}\to\frac{2r}{na_0}=\rho$. The remaining bra-ket $\langle r_x{=}0,r_y{=}0,r_z{=}0|\phi_{\lambda{=}n}\rangle$ is computed in Eq.~(\ref{eq: norm_coord}).
We compute the overall normalization 
by employing the results from Eq.~(\ref{eq: wf_norm}). After some significant algebra, we find that the normalized radial wavefunction satisfies
\begin{equation}
R_{nl}(\rho)=\sqrt{\left (\frac{2}{na_0}\right )^3\frac{(n-l-1)!}{2n(n+l)!}}\,\rho^lL^{2l+1}_{n-l-1}(\rho)e^{-\frac{1}{2}\rho},
\end{equation}
which is the standard result for Hydrogen. 

\section{The wavefunction in momentum space}

Unfortunately, the derivation of the wavefunction in momentum space is even more complicated. This is because neither the coordinate eigenfunctions, nor the momentum eigenfunctions are eigenstates of the radial momentum operator. Indeed, even though this operator is manifestly Hermitian (or perhaps more correct technically, symmetric), it is not self adjoint~\cite{radial_momentum}. Hence, it has no eigenfunctions that we can employ to expand the momentum wavefunctions with respect to. Nevertheless, one can still determine the momentum wavefunctions in a straightforward fashion, employing only algebraic methods (we do require some identities of binomial coefficients, which we could only derive using calculus, as we illustrate below). The approach we develop is the most direct way to obtain the momentum wavefunctions and does not require any complex Fourier transformations.

Unfortunately, these technical issues require that the journey to compute the momentum wavefunctions is a long one. The momentum wavefunction for the general case is given by the following normalized form:
\begin{align}
\psi_{nl}(p_x,p_y,p_z)&=\frac{(\sqrt{2a_0})^{n-l-1}}{e^{n-l-1}\sqrt{\prod_{j=l+1}^{n-1}\left (\frac{1}{j^2}-\frac{1}{n^2}\right )}}\nonumber\\
&\times\langle p_x,p_y,p_z|\hat B_r^\dagger(l)\cdots\hat B_r^\dagger(n-2)\hat r^{n-l-1}P_h^l(\hat r_x,\hat r_y,\hat r_z)|\phi_{\lambda{=}n}\rangle.
\label{eq: momentum_wf}
\end{align}

Because neither the momentum eigenstate nor the auxiliary Hamiltonian ground state are  eigenstates of the radial coordinate or the radial momentum, we cannot immediately evaluate the matrix element that defines the momentum wavefunction. Instead, our strategy is to first determine how one evaluates the expectation value of a power of the radial coordinate (we will actually find it simpler to find the matrix element of a reverse Bessel polynomial of the radial coordinate, as we develop below). Then we show how one removes the harmonic polynomial from the matrix element. Finally, we put both together to evaluate the expectation value of the full operator. Each step is rather long and technical.

Our first step is to try to evaluate the matrix element of powers of $\hat r$ between the momentum eigenstate and the auxiliary Hamiltonian ground state, which we define to be
\begin{equation}
    \mu_m(\vec{p},\lambda{=}n)=\langle p_x,p_y,p_z|\hat r^m|\phi_{\lambda{=}n}\rangle.
\end{equation}

Since we cannot evaluate the $\hat r$ operator on the momentum eigenstate or the auxiliary Hamiltonian ground state, we rewrite the $\hat r$ term as $\sum_\alpha \hat r_\alpha^2/\hat r$ and then we can use the fact that each of the Cartesian annihilation operators with $\lambda{=}n$ annihilate the $|\phi_{\lambda{=}n}\rangle$ state to find that
\begin{align}
     \mu_m(\vec{p},\lambda{=}n)&=
    \langle p_x,p_y,p_z|\hat r^{m-1}\sum_\alpha \hat r_\alpha \frac{\hat r_\alpha}{\hat r}|\phi_{\lambda{=}n}\rangle\nonumber\\
    &=\sum_\alpha \frac{na_0}{i\hbar}
    \langle p_x,p_y,p_z|\hat r^{m-1}\hat r_\alpha\hat p_\alpha|\phi_{\lambda{=}n}\rangle\nonumber\\
    &=\sum_\alpha \frac{na_0}{i\hbar}
    \langle p_x,p_y,p_z|\hat p_\alpha\hat r^{m-1}\hat r_\alpha |\phi_{\lambda{=}n}\rangle+\sum_\alpha \frac{na_0}{i\hbar}
    \langle p_x,p_y,p_z|[\hat r^{m-1}\hat r_\alpha,\hat p_\alpha] |\phi_{\lambda{=}n}\rangle\nonumber\\
    &=\sum_\alpha \frac{na_0}{i\hbar}
    p_\alpha\langle p_x,p_y,p_z|\hat r^{m-1}\hat r_\alpha |\phi_{\lambda{=}n}\rangle\nonumber\\
    &+\sum_\alpha na_0(m-1)
    \langle p_x,p_y,p_z|\hat r^{m-3}\hat r_\alpha^2 |\phi_{\lambda{=}n}\rangle+3na_0\langle p_x,p_y,p_z|\hat r^{m-1} |\phi_{\lambda{=}n}\rangle\nonumber\\
    &=na_0(m+2)\mu_{m-1}(\vec{p},\lambda{=}n)+\sum_\alpha\left (\frac{na_0}{i\hbar}\right )^2p_\alpha\langle p_x,p_y,p_z|\hat r^m\hat p_\alpha|\phi_{\lambda{=}n}\rangle\nonumber\\
    &=\left (\frac{na_0p}{i\hbar}\right )^2\mu_m(\vec{p},\lambda{=}n)+na_0(m+2)\mu_{m-1}(\vec{p},\lambda{=}n)\nonumber\\
    &+\frac{(na_0)^2}{i\hbar}\sum_\alpha mp_\alpha \langle p_x,p_y,p_z|\hat r^{m-2}\hat r_\alpha|\phi_{\lambda{=}n}\rangle.
\end{align}
In this derivation, we used the annihilation operator relation to create a momentum operator on the right, commute it through to the left, where it can operate on the bra, collect the commutator terms, and repeat with the remaining $\hat r_\alpha$ term;  note that we also used the fact that $p^2=p_x^2+p_y^2+p_z^2$. We combine the $\mu_m(\vec{p},\lambda{=}n)$ terms on the left and then repeat the procedure to remove additional powers of $\hat r$ as follows:
\begin{align}
&\left [ 1+\left (\frac{na_0p}{\hbar}\right )^2\right ]\mu_m(\vec{p},\lambda{=}n)\nonumber\\
&~=na_0(m+2)\mu_{m-1}(\vec{p},\lambda{=}n)+\frac{(na_0)^3}{(i\hbar)^2} \sum_\alpha mp_\alpha\langle p_x,p_y,p_z|\hat r^{m-1}\hat p_\alpha|\phi_{\lambda{=}n}\rangle\nonumber\\
&~=na_0(m+2)\mu_{m-1}(\vec{p},\lambda{=}n)+
\left (\frac{na_0p}{i\hbar}\right )^2na_0m\mu_{m-1}(\vec{p},\lambda{=}n)\nonumber\\
&~~~~~~+\frac{na_0}{i\hbar}(na_0)^2m(m-1)\sum_\alpha p_\alpha\langle p_x,p_y,p_z|\hat r^{m-3}\hat r_\alpha|\phi_{\lambda{=}n}\rangle\nonumber\\
&~=na_0(m+2)\mu_{m-1}(\vec{p},\lambda{=}n)-\left (\frac{na_0p}{\hbar}\right )^2\sum_{k=1}^{m}\frac{m!}{(m-k)!}(na_0)^k\mu_{m-k}(\vec{p},\lambda{=}n).
\label{eq: r_m_multi_recur}
\end{align}
The last line results by repeating the above procedure $m-1$ more times.
This is an $m$-term recurrence relation, which is not simple to work with. We can reduce it to a two-term recurrence relation by taking the recurrence relation for $m-1$ and multiplying it by $mna_0$. This yields
\begin{align}
mna_0\left [ 1+\left (\frac{na_0p}{\hbar}\right )^2\right ]\mu_{m-1}(\vec{p},\lambda{=}n)&=(na_0)^2m(m+1)\mu_{m-2}(\vec{p},\lambda{=}n)\nonumber\\
&-\left (\frac{na_0p}{\hbar}\right )^2\sum_{k=2}^m\frac{m!}{(m-k)!}(na_0)^k\mu_{m-k}(\vec{p},\lambda{=}n).
\end{align}
We use this result to replace the $m-1$ terms in the summation in Eq.~(\ref{eq: r_m_multi_recur}) with $2\le k\le m$, to find that
\begin{equation}
\left [ 1+\left (\frac{na_0p}{\hbar}\right )^2\right ]\mu_m(\vec{p},\lambda{=}n)=2na_0(m+1)\mu_{m-1}(\vec{p},\lambda{=}n)-(na_0)^2m(m+1)\mu_{m-2}(\vec{p},\lambda{=}n)
\end{equation}
or
\begin{equation}
\mu_m(\vec{p},\lambda{=}n)=\frac{2na_0(m+1)}{1+\left (\frac{na_0p}{\hbar}\right )^2}\mu_{m-1}(\vec{p},\lambda{=}n)-\frac{(na_0)^2m(m+1)}{1+\left (\frac{na_0p}{\hbar}\right )^2}\mu_{m-2}(\vec{p},\lambda{=}n).
\end{equation}

Our next step is to solve the recurrence relation. We define the $m$th moment matrix element via
\begin{equation}
    \mu_m(\vec{p},\lambda{=}n)=\frac{Q_m(\xi_n^2)(na_0)^m}{\left [ 1+\xi_n^2\right ]^m}\langle p_x,p_y,p_z|\phi_{\lambda{=}n}\rangle
    \label{eq: mu_final}
\end{equation}
where $Q_m(\xi_n^2)$ is a $m$-degree polynomial in $\xi_n^2=(na_0p/\hbar)^2$. This recurrence relation then becomes
\begin{equation}
    Q_m(\xi_n^2)=2(m+1)Q_{m-1}(\xi_n^2)-m(m+1)(1+\xi_n^2)Q_{m-2}(\xi_n^2).
\end{equation}
This recurrence relation is solved by
\begin{equation}
    Q_m(\xi_n^2)=\frac{(m+2)!}{2}\sum_{j=0}^{\left \lfloor \frac{m+1}{2}\right \rfloor}(-1)^j\frac{(m+1)!}{(m+1-2j)!(2j+1)!}\xi_n^{2j}.
    \label{eq: q_final_sum}
\end{equation}
To verify this, we must go through the inductive proof for even and odd $m$.

To start, we need to establish the base cases. The first one, is for $m=0$, which is trivial, because $Q_0(\xi_n^2)=1$. The case with $m=1$ requires some work. In particular, we have
\begin{equation}
\langle p_x,p_y,p_z|\hat r|\phi_{\lambda{=}n}\rangle=\sum_\alpha\langle p_x,p_y,p_z|\hat r_\alpha\frac{\hat r_\alpha}{\hat r}|\phi_{\lambda{=}n}\rangle.
\end{equation}
Using the fact that $\hat A_\alpha(\lambda{=}n)|\phi_{\lambda{=}n}\rangle=0$, then yields
\begin{equation}
\langle p_x,p_y,p_z|\hat r|\phi_{\lambda{=}n}\rangle=\frac{na_0}{i\hbar}\sum_\alpha\langle p_x,p_y,p_z|\hat r_\alpha \hat p_\alpha|\phi_{\lambda{=}n}\rangle.
\end{equation}
We move the $\hat p_\alpha$ operator to the left. This gives
\begin{equation}
    \langle p_x,p_y,p_z|\hat r|\phi_{\lambda{=}n}\rangle=\frac{na_0}{i\hbar}\sum_\alpha p_\alpha\langle p_x,p_y,p_z|\hat r_\alpha |\phi_{\lambda{=}n}\rangle+3na_0\langle p_x,p_y,p_z|\phi_{\lambda{=}n}\rangle.
\end{equation}
We insert $1=\hat r/\hat r$ in the center of the first term on the right, and use the annihilation operator relation again
\begin{equation}
    \langle p_x,p_y,p_z|\hat r|\phi_{\lambda{=}n}\rangle=\left (\frac{na_0}{i\hbar}\right )^2\sum_\alpha p_\alpha\langle p_x,p_y,p_z|\hat r\hat p_\alpha |\phi_{\lambda{=}n}\rangle+3na_0\langle p_x,p_y,p_z|\phi_{\lambda{=}n}\rangle.
\end{equation}
Once more, we move the $\hat p_\alpha$ to the left.
\begin{align}
    \langle p_x,p_y,p_z|\hat r|\phi_{\lambda{=}n}\rangle&=-\left (\frac{na_0p}{\hbar}\right )^2 \langle p_x,p_y,p_z|\hat r |\phi_{\lambda{=}n}\rangle+\frac{(na_0)^2}{i\hbar}\sum_\alpha p_\alpha\langle p_x,p_y,p_z|\frac{\hat r_\alpha}{\hat r}|\phi_{\lambda{=}n}\rangle\nonumber\\
    &+3na_0\langle p_x,p_y,p_z|\phi_{\lambda{=}n}\rangle.
\end{align}
The operator in the middle term on the right can be replaced by $\hat p_\alpha$. Then we collect terms and solve the the first-order term. We end up with
\begin{equation}
    \langle p_x,p_y,p_z|\hat r|\phi_{\lambda{=}n}\rangle=\frac{3-\left (\frac{na_0p}{\hbar}\right )^2}{1+\left (\frac{na_0p}{\hbar}\right )^2}(na_0)\langle p_x,p_y,p_z|\phi_{\lambda{=}n}\rangle,
\end{equation}
or $Q_1(\xi_n^2)=3-\xi_n^2$.

Using the recurrence relation, then gives
\begin{equation}
    Q_2(\xi_n^2)=6(3-\xi_n^2)-6(1-\xi_n^2)=12(1-\xi_n^2),
\end{equation}
which is exactly what we get if we substitute $m=2$ into Eq.~(\ref{eq: q_final_sum}). Of course this is also what one would get if one derived the polynomial directly using the operator techniques discussed above, but this approach becomes tedious for large $m$. Instead, we now show the proof by induction that establishes these results for all $m$.

First we work with even $m=2k$, assuming the relation holds for all cases up to $m-1$. Then we find that the recurrence relation becomes the following, after replacing $Q_{m-1}$ and $Q_{m-2}$ by their summation forms, 
\begin{align}
Q_{2k}(\xi_n^2)&=2(2k+1)\frac{(2k+1)!}{2}\sum_{j=0}^k\frac{(2k)!(-1)^j}{(2k-2j)!(2j+1)!}\xi_n^{2j}\nonumber\\
&-2k(2k+1)(1+\xi_n^2)\frac{(2k)!}{2}\sum_{j=0}^{k-1}\frac{(2k-1)!(-1)^j}{(2k-2j-1)!(2j+1)!}\xi_n^{2j}.
\end{align}
We examine the coefficients of $\xi_n^{2j}$ for $0\le j\le k$ to determine the polynomial for $m=2k$. There are three different cases. For $j=0$, we have
\begin{equation}
\frac{(2k+1)!(2k+1)!}{(2k)!}-\frac{(2k)!(2k+1)!}{2(2k-1)!}=(2k+1)!(k+1)=\frac{(2k+2)!}{2},
\end{equation}
which is the $j=0$ term in Eq.~(\ref{eq: q_final_sum}). Next, we examine the case with $1\le j<k$:
\begin{align}
&\frac{(2k+1)!(2k+1)!(-1)^j}{(2k-2j)!(2j+1)!}-\frac{(2k)!(2k+1)!(-1)^j}{2(2k-2j-1)!(2j+1)!}+
\frac{(2k)!(2k+1)!(-1)^j}{2(2k-2j+1)!(2j-1)!}\nonumber\\
&=\frac{(2k)!(2k+1)!(-1)^j}{2(2k-2j+1)!(2j+1)!}
[2(2k+1)(2k-2j+1-(2k-2j)(2k-2j+1)+2j(2j+1)]\nonumber\\
&=\frac{(2k+1)!(2k+2)!(-1)^j}{2(2k+1-2j)!(2j+1)!},
\end{align}
which also agrees with the coefficient of the $j$th term in the summation in Eq.~(\ref{eq: q_final_sum}). The last term we need to check is
for $j=k$, which is
\begin{equation}
(2k+1)!(-1)^k+(2k+1)!(-1)^kk=\frac{(2k+2)!(-1)^k}{2},
\end{equation}
which also agrees with the $j=k$ term in Eq.~(\ref{eq: q_final_sum}). This establishes the inductive proof for the case of even $m$. 

Next, we work out the case with odd $m=2k+1$. Once again, we start with the two-term recurrence relation and replace the polynomials $Q_{2k}$ and $Q_{2k-1}$ by their corresponding power series expressions, which are assumed to be true. This yields
\begin{align}
Q_{2k+1}(\xi_n^2)&=2(2k+2)\sum_{j=0}^{k}\frac{(2k+2)!(2k+1)!(-1)^j}{2(2k+1-2j)!(2j+1)!}\xi_n^{2j}\nonumber\\
&-(2k+1)(2k+2)(1+\xi_n^2)\sum_{j=0}^k\frac{(2k+1)!(2k)!(-1)^j}{2(2k-2j)!(2j+1)!}\xi_n^{2j}.
\end{align}
We examine the coefficients of $\xi_n^{2j}$. There are three cases: (i) $j=0$; (ii) $1\le j\le k$; and (iii) $j=k+1$. For the first case, we have
\begin{equation}
    \frac{(2k+2)!(2k+2)!}{(2k+1)!}-\frac{(2k+1)!(2k+2)!}{2(2k)!}=(2k+2)!\left [(2k+2)-\left (k+\frac{1}{2}\right )\right ]=\frac{(2k+3)!}{2},
\end{equation}
which is the $j=0$ term in Eq.~(\ref{eq: q_final_sum}) for $m=2k+1$. The general case, with $1\le j\le k$ has
\begin{align}
&\frac{(2k+2)!(2k+2)!(-1)^j}{(2k+1-2j)!(2j+1)!}-\frac{(2k+1)!(2k+2)!(-1)^j}{2(2k-2j)!(2j+1)!}+\frac{(2k+1)!(2k+2)!(-1)^j}{2(2k-2j+2)!(2j-1)!}\nonumber\\
    &~=\frac{(2k+1)!(2k+2)!(-1)^j}{2(2k-2j+2)!(2j+1)!}[2(2k+2)(2k-2j+2)\nonumber\\
    &~-(2k-2j+1)(2k-2j+2)+2j(2j+1)]\nonumber\\
    &~=\frac{(2k+3)!}{2}\frac{(2k+2)!(-1)^j}{(2k+2-2j)!(2j+1)!},
\end{align}
which agrees with the $j$th term ( $1\le j\le k$) in Eq.~(\ref{eq: q_final_sum}) for $m=2k+1$.

The final case is $j=k+1$, where we have
\begin{equation}
(-1)^{k+1}\frac{(2k+1)!(2k+2)!}{2(2k+1)!}=\frac{(2k+2)!}{2}(-1)^{k+1},
\end{equation}
which also is the coefficient of the $j=k+1$ term in Eq.~(\ref{eq: q_final_sum}) for $m=2k+1$.

With the proof by induction complete, we have established Eq.~(\ref{eq: mu_final}) with the polynomials $Q_m(\xi_n^2)$ determined by Eq.~(\ref{eq: q_final_sum}). We are now ready to move on to the second phase, which shows how to extract the harmonic polynomial from the matrix element. This step is also technical and complex.

We begin with the task of evaluating the following matrix element
\begin{equation}
    \langle p_x,p_y,p_z|\hat r^{m-l}P_h^l(\hat r_x,\hat r_y, \hat r_z)|\phi_{\lambda{=}n}\rangle.
    \label{eq: power_harm_poly}
\end{equation}
Similar to the previous calculations, we cannot evaluate any of these operators against the momentum states, but we can exchange $\hat r_\alpha$ terms by $na_0\hat r\hat p_\alpha/(i\hbar)$ when it acts on the state $|\phi_{\lambda{=}n}\rangle$. Then the $p_\alpha$ term can be moved to the left, where it can operate on the momentum eigenstate, which is an eigenvector for that operator. Because it does not commute with the harmonic polynomial or the radial coordinate, we have to evaluate some additional commutators. This is shown in the next equation
\begin{align}
&\langle p_x,p_y,p_z|\hat r^{m-l}P_h^l(\hat r_x,\hat r_y, \hat r_z)|\phi_{\lambda{=}n}\rangle=-\frac{1}{i\hbar l}
\langle p_x,p_y,p_z|\hat r^{m-l}\sum_\alpha \left [\hat p_\alpha,P_h^l(\hat r_x,\hat r_y,\hat r_z)\right ]\hat r_\alpha|\phi_{\lambda{=}n}\rangle\nonumber\\
&~=-\frac{na_0}{(i\hbar)^2 l}
\sum_\alpha\langle p_x,p_y,p_z|\hat r^{m-l} \left [\hat p_\alpha,P_h^l(\hat r_x,\hat r_y,\hat r_z)\right ]\hat r\hat p_\alpha|\phi_{\lambda{=}n}\rangle\nonumber\\
&~=-\frac{na_0}{(i\hbar)^2 l}
\sum_\alpha p_\alpha\langle p_x,p_y,p_z|\hat r^{m+1-l} \left [\hat p_\alpha,P_h^l(\hat r_x,\hat r_y,\hat r_z)\right ] |\phi_{\lambda{=}n}\rangle\nonumber\\
&~~~-\frac{na_0}{(i\hbar)^2 l}
\sum_\alpha\langle p_x,p_y,p_z|\left [ \hat r^{m+1-l},\hat p_\alpha \right ] \left [\hat p_\alpha,P_h^l(\hat r_x,\hat r_y,\hat r_z)\right ]|\phi_{\lambda{=}n}\rangle\nonumber\\
&~~~+\frac{na_0}{(i\hbar)^2 l}
\sum_\alpha\langle p_x,p_y,p_z|\hat r^{m+1-l}\left[\hat p_\alpha,\left [\hat p_\alpha,P_h^l(\hat r_x,\hat r_y,\hat r_z)\right ]\right ]|\phi_{\lambda{=}n}\rangle\nonumber\\
&~=-\frac{na_0}{(i\hbar)^2 l}
\sum_\alpha p_\alpha\langle p_x,p_y,p_z|\hat r^{m+1-l} \left [\hat p_\alpha,P_h^l(\hat r_x,\hat r_y,\hat r_z)\right ] |\phi_{\lambda{=}n}\rangle\nonumber\\
&~~~-\frac{na_0(m+1-l)}{i\hbar l}
\sum_\alpha\langle p_x,p_y,p_z|\hat r^{m-l} \left [\hat p_\alpha,P_h^l(\hat r_x,\hat r_y,\hat r_z)\right ]\frac{\hat r_\alpha}{\hat r}|\phi_{\lambda{=}n}\rangle\nonumber\\
&~=-\frac{na_0}{(i\hbar)^2 l}
\sum_\alpha p_\alpha\langle p_x,p_y,p_z|\hat r^{m+1-l} \left [\hat p_\alpha,P_h^l(\hat r_x,\hat r_y,\hat r_z)\right ] |\phi_{\lambda{=}n}\rangle\nonumber\\
&~~~-\frac{(na_0)^2(m+1-l)}{(i\hbar)^2 l}
\sum_\alpha\langle p_x,p_y,p_z|\hat r^{m-l} \left [\hat p_\alpha,P_h^l(\hat r_x,\hat r_y,\hat r_z)\right ]\hat p_\alpha|\phi_{\lambda{=}n}\rangle\nonumber\\
&~=-\frac{na_0}{(i\hbar)^2 l}
\sum_\alpha p_\alpha\langle p_x,p_y,p_z|\hat r^{m-l} [\hat r+na_0(m-l+1)]\left [\hat p_\alpha,P_h^l(\hat r_x,\hat r_y,\hat r_z)\right ] |\phi_{\lambda{=}n}\rangle\nonumber\\
&~~~+(na_0)^2(m-l+1)(m-l)
\langle p_x,p_y,p_z|\hat r^{m-l-2} P_h^l(\hat r_x,\hat r_y,\hat r_z)|\phi_{\lambda{=}n}\rangle.
\end{align}
Let's take a moment to understand what this last equation has accomplished. The first term has replaced one occurrence of $\hat r_\alpha$ in the harmonic polynomial by $p_\alpha$; the power of the radial coordinate was multiplied by a monomial in $\hat r$, while the remaining harmonic polynomial of the position operators has had its order reduced by one. The second term reduced the power of the radial coordinate operator by two, leaving the harmonic polynomial unchanged.

This recursion is complex, so we first solve it for the simplest case where $m=l=n-1$, which corresponds to the $\psi_{nn-1}(\vec{p})$ set of wavefunctions, which have maximal angular momentum for the given principal quantum number. In this case, after $j$ iterations ($j<n$), we have
\begin{align}
&\langle p_x,p_y,p_z|P_h^{n-1}(\hat r_x,\hat r_y,\hat r_z)|\phi_{\lambda{=}n}\rangle=\frac{(na_0)^{2j}(n-1-j)!}{\hbar^{2j}(n-1)!}
\sum_{\alpha_1\cdots\alpha_j}p_{\alpha_1}\cdots p_{\alpha_j}\nonumber\\
&~~~~~~~~~~\times\langle p_x,p_y,p_z| \theta_{j}\left (\frac{\hat r}{na_0}\right )\left [\hat p_{\alpha_j},\cdots \left [\hat p_{\alpha_1},P_h^{n-1}(\hat r_x,\hat r_y,\hat r_z)\right ]\cdots\right ]_j|\phi_{\lambda{=}n}\rangle,
\end{align}
where $\theta_0=1$ and the $j$ subscript on the commutators is to remind us that there are $j$ nested commutators. We remove another power of $\hat r_\alpha$ from the harmonic polynomial to determine the recurrence relation of the polynomial $\theta$. This uses similar steps as we employed above, and after bringing one power of $\hat p_{\alpha_{j+1}}$ all the way to the left and using the fact that $[f(\hat r),\hat p_\alpha]=[f(\hat r),\hat p_r]\hat r_\alpha/\hat r$, we obtain
\begin{align}
&\langle p_x,p_y,p_z|P_h^{n-1}(\hat r_x,\hat r_y,\hat r_z)|\phi_{\lambda{=}n}\rangle\nonumber\\
&~~~=-\frac{(na_0)^{2j}(n-2-j)!}{i\hbar^{2j+1}(n-1)!}
\sum_{\alpha_1\cdots\alpha_{j+1}}p_{\alpha_1}\cdots p_{\alpha_j} \nonumber\\
&~~~\times\langle p_x,p_y,p_z|\hat r \theta_{j}\left (\frac{\hat r}{na_0}\right )\left [\hat p_{\alpha_{j+1}},\cdots \left [\hat p_{\alpha_1},P_h^{n-1}(\hat r_x,\hat r_y,\hat r_z)\right ]\cdots\right ]_{j+1}\frac{\hat r_{\alpha_{j+1}}}{\hat r}|\phi_{\lambda{=}n}\rangle\nonumber\\
&~=\frac{(na_0)^{2j+2}(n-2-j)!}{\hbar^{2j+2}(n-1)!}
\Biggr \{\sum_{\alpha_1\cdots\alpha_{j+1}}p_{\alpha_1}\cdots p_{\alpha_{j+1}}\nonumber\\&~~~\times\langle p_x,p_y,p_z| 
\frac{\hat r}{na_0}\theta_{j}\left (\frac{\hat r}{na_0}\right )\left [\hat p_{\alpha_{j+1}},\cdots \left [\hat p_{\alpha_1},P_h^{n-1}(\hat r_x,\hat r_y,\hat r_z)\right ]\cdots\right ]_{j+1}|\phi_{\lambda{=}n}\rangle\nonumber\\
&~~~+\sum_{\alpha_1\cdots\alpha_{j+1}}p_{\alpha_1}\cdots p_{\alpha_{j}}\nonumber\\
&~~~\times\langle p_x,p_y,p_z|\left [\frac{\hat r}{na_0}\theta_{j}\left (\frac{\hat r}{na_0}\right ),\hat p_r\right ] \left [\hat p_{\alpha_{j+1}},\cdots \left [\hat p_{\alpha_1},P_h^{n-1}(\hat r_x,\hat r_y,\hat r_z)\right ]\cdots\right ]_{j+1}\frac{\hat r_{\alpha_{j+1}}}{\hat r}|\phi_{\lambda{=}n}\rangle\Biggr \}\nonumber\\
\label{eq: theta_gen_rec}
\end{align}
Since $\theta_j$ is an order $j$ polynomial, we can continue the recursion above $j$ more times, to end with
\begin{align}
\theta_{j+1}\left (\frac{\hat r}{na_0}\right )
&=\frac{\hat r}{na_0}\theta_j\left (\frac{\hat r}{na_0}\right )+\frac{na_0}{i\hbar}\left [\frac{\hat r}{na_0}\theta_j\left (\frac{\hat r}{na_0}\right ),\hat p_r\right ]+\cdots\nonumber\\
&+\left (\frac{na_0}{i\hbar}\right )^{j+1}\left [\cdots \left [\frac{\hat r}{na_0}\theta\left (\frac{\hat r}{na_0}\right ),\hat p_r\right ],\cdots \hat p_r\right ]_{j+1}\nonumber\\
&=\sum_{k=0}^{j+1}\left (\frac{na_0}{i\hbar}\right )^k\left [ \cdots \left [\frac{\hat r}{na_0}\theta_j\left (\frac{\hat r}{na_0}\right ),\hat p_r\right ], \cdots \hat p_r\right ]_k,
\label{eq: theta_commutator}
\end{align}
where the subscript $k$ means we have $k$ iterated commutators with $\hat p_r$.

The wavefunction then becomes
\begin{align}
&\psi_{nn-1}(\vec{p})=\frac{(na_0)^{2n-2}}{\hbar^{2n-2}(n-1)!}\sum_{\alpha_1\cdots\alpha_{n-1}}
p_{\alpha_1}\cdots p_{\alpha_{n-1}}\nonumber\\
&~\times\langle p_x,p_y,p_z|\theta_{n-1}\left ( \frac{\hat r}{na_0}\right )\left [ \hat p_{\alpha_{n-1}},\cdots\left [\hat p_{\alpha_1},P_h^{n-1}(\hat r_x,\hat r_y,\hat r_z)\right ]\cdots\right ]_{n-1}|\phi_{\lambda{=}n}\rangle\nonumber\\
&~=\frac{(na_0)^{2n-2}}{(i\hbar)^{n-1}}P_h^{n-1}(p_x,p_y,p_z)\langle p_x,p_y,p_z|\theta_{n-1}\left (\frac{\hat r}{na_0}\right )|\phi_{\lambda{=}n},\rangle\nonumber\\
\end{align}
which requires determining the expectation value of the polynomial. Note that the emergence of the harmonic polynomial as a function of the momenta is subtle. The multiple commutators sequentially remove every factor of $\hat r_\alpha$ in each term of the polynomial leaving behind the constant prefactor. This is then multiplied by the corresponding $p_\alpha$ factors, which allow us to reconstruct the harmonic polynomial as a function of momentum as follows:
\begin{equation}
    P_h^{n-1}(p_x,p_y,p_z)=\sum_{\alpha_1\cdots \alpha_{n-1}}p_{\alpha_1}\cdots p_{\alpha_{n-1}}\frac{[\hat p_{\alpha_{n-1}},[\cdots ,[\hat p_1,P_h^{n-1}(\hat r_x,\hat r_y,\hat r_z)]\cdots ]_{n-1}}{(-i\hbar)^{n-1}(n-1)!}.
\end{equation}
So, we have only the matrix element left to evaluate. Since we have already found the matrix elements for arbitrary powers of $\hat r$, we can then determine this result once we have an explicit formula for the $\theta_{n-1}$ polynomial. 

The polynomial recurrence relation can be rewritten in terms of derivatives as follows:
\begin{equation}
    \theta_{j}\left ( \frac{r}{na_0}\right )=\sum_{k=0}^{j}(na_0)^{k}\frac{d^k}{dr^k}\left [\frac{r}{na_0}\theta_{j-1}\left ( \frac{r}{na_0}\right )\right ].
    \label{eq: bessel_deriv}
\end{equation}
This polynomial is a reverse Bessel polynomial~\cite{bessel_poly_book}, which satisfies
\begin{equation}
    \theta_j\left ( \frac{r}{na_0}\right )=\sum_{k=0}^j\frac{(j+k)!}{(j-k)!k!}\frac{1}{2^k}\left (\frac{r}{na_0}\right )^{j-k}.
    \label{eq: theta_deriv}
\end{equation}
The proof can be done with or without calculus---it is identical, because $1/(i\hbar)$ times the commutator of the radial momentum operator with a power of the radial coordinate operator is the same as the derivative with respect to the radial coordinate operator. The base case has $\theta_0=1$ and $\theta_1=\hat r/na_0+1$ from Eq.~(\ref{eq: theta_commutator}). We next work with the derivative form and note that
\begin{equation}
    (na_0)^{k}\frac{d^k}{dr^k}\left [\frac{r}{na_0}\theta_{j-1}\left (\frac{r}{na_0}\right )\right ]=\sum_{k'=0}^{j-k}\frac{(j-1+k')!(j-k')}{(j-k'-k)!k'!}\frac{1}{2^{k'}}\left ( \frac{r}{na_0}\right )^{j-k'-k}.
\end{equation}
Plugging this into the derivative relation in Eq.~(\ref{eq: bessel_deriv}) yields
\begin{equation}
 \theta_j\left ( \frac{r}{na_0}\right )=
 \sum_{k=0}^{j}\sum_{k'=0}^{j-k}\frac{(j-1+k')!(j-k')}{(j-k'-k)!k'!}\frac{1}{2^{k'}}
    \left (\frac{r}{na_0}\right )^{j-k'-k}
\end{equation}

We now change the summation indices. We let $m=k+k'$ and $m'=k'$. Then we have
\begin{equation}
 \theta_j\left ( \frac{r}{na_0}\right )=
 \sum_{m=0}^{j}\sum_{m'=0}^{m}\frac{(j-1+m')!(j-m')}{(j-m)!m'!}\frac{1}{2^{m'}}
    \left (\frac{r}{na_0}\right )^{j-m}.
\end{equation}
One can easily verify the sum over $m'$ by induction (for fixed $j$) to show that
\begin{equation}
    \sum_{m'=0}^m\frac{(j-1+m')!(j-m')}{m'!2^{m'}}=\frac{(j+m)!}{m!2^m}.
\end{equation}
We easily see the base case $m=0$ is satisfied since both sides are $j!$. Assume it holds up to $m$, then after separating the sum for $m'$ up to $m$ and adding in the term with $m'=m+1$, we have
\begin{align}
\sum_{m'=0}^{m+1}\frac{(j-1+m')!(j-m')}{m'!2^{m'}}&=\frac{(j+m)!}{m!2^m}+\frac{(j+m)!(j-m-1)}{(m+1)!2^{m+1}}\nonumber\\
&=\frac{(j+m)!}{m!2^m}\left (1+\frac{j-m-1}{2m+2}\right )=\frac{(j+m+1)!}{(m+1)!2^{m+1}}.
\end{align}
So, the polynomial becomes
\begin{equation}
     \theta_j\left ( \frac{r}{na_0}\right )=
 \sum_{m=0}^{j}\frac{(j+m)!}{(j-m)!m!}\frac{1}{2^{m}}
   \left (\frac{r}{na_0}\right )^{j-m},
   \label{eq: rev_bessel_poly_def}
\end{equation}
which establishes that $\theta_j$ is the $j$th inverse Bessel polynomial. Note that the recurrence relation with derivatives in Eq.~(\ref{eq: bessel_deriv}) appears to be a new relation for the reverse Bessel polynomials (that is, it is not in Ref.~\cite{bessel_poly_book}).

We now must return to the problem of determining the matrix element in Eq.~(\ref{eq: power_harm_poly}). We already have already found that the result for $m=l=n-1$ can be expressed in terms of the reverse Bessel polynomial $\theta_{n-1}$. This result can be immediately generalized to $m=l-1$, which can be expressed in terms of a reverse Bessel polynomial $\theta_{l-1}$.
To be more concrete consider the following:
\begin{align}
\langle p_x,p_y,p_z|\frac{1}{\hat r}P_h^l(\hat r_x,\hat r_y,\hat r_z)|\phi_{\lambda{=}n}\rangle&=\frac{i}{\hbar l}\sum_\alpha \langle p_x,p_y,p_z|[\hat p_\alpha,P_h^l(\hat r_x,\hat r_y,\hat r_z)]\frac{\hat r_\alpha}{\hat r}|\phi_{\lambda{=}n}\rangle\nonumber\\
&=\frac{na_0}{\hbar^2l}\sum_\alpha p_\alpha\langle p_x,p_y,p_z|[\hat p_\alpha,P_h^l(\hat r_x,r_y,r_z)]|\phi_{\lambda{=}n}\rangle\nonumber\\
&=\frac{(na_0)^{2l-1}}{(i\hbar)^l}P_h^l(p_x,p_y,p_z)\langle p_x,p_y,p_z|\theta_{l-1}\left (\frac{\hat r}{na_0}\right )|\phi_{\lambda{=}n}\rangle.\nonumber\\
\end{align}

\begin{figure}
    \centering
    \includegraphics[width=0.55\textwidth]{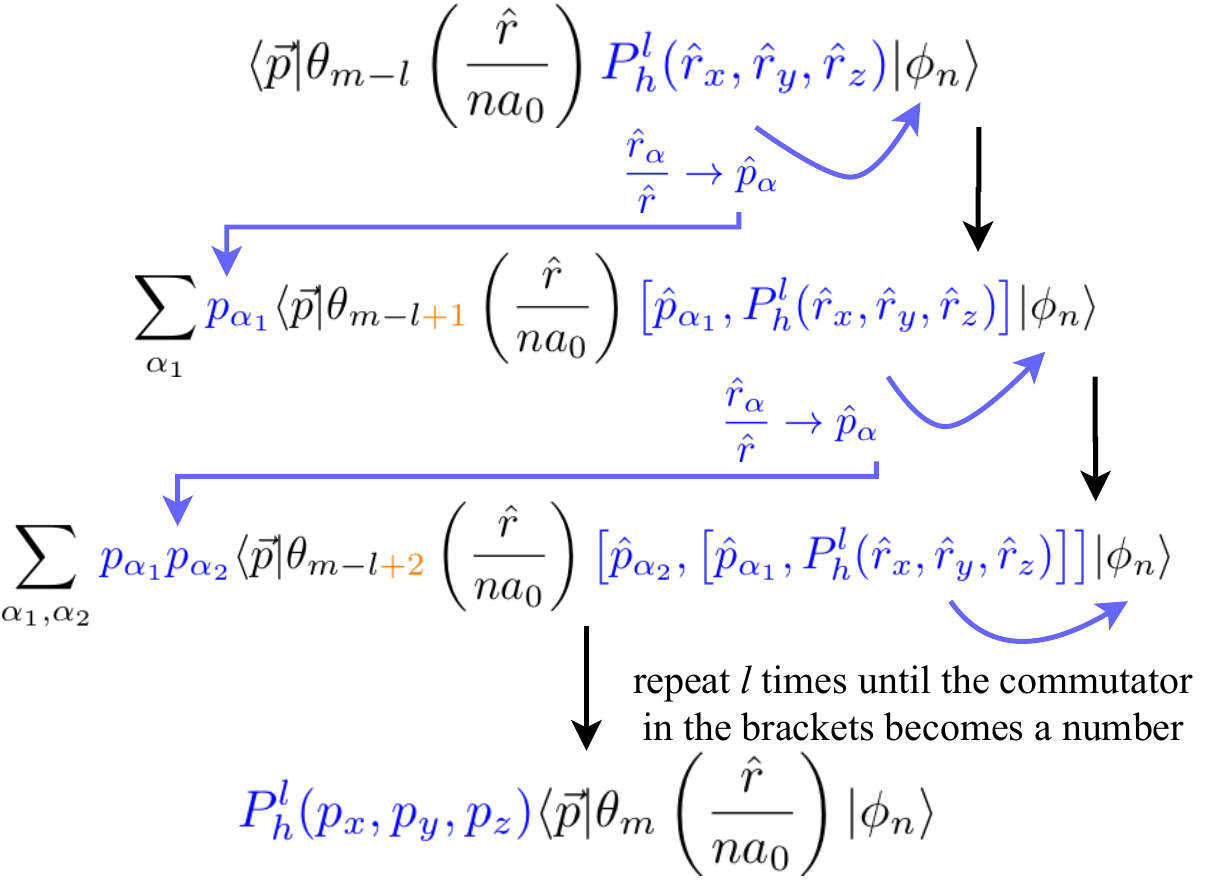}
    \caption{Schematic for how to remove the harmonic polynomial from the operator matrix element to determine the wavefunction in momentum space. Here $\theta$ is a reverse Bessel polynomial and $P_h^l$ is a harmonic polynomial.}
    \label{fig:bessel}
\end{figure}

Unfortunately, this approach does not produce simple results for other powers. What does produce simple results is matrix elements of reverse Bessel polynomials times harmonic polynomials. We now describe why. As we already saw in Eq.~(\ref{eq: theta_gen_rec}), if we have an expectation value of a reverse Bessel polynomial in $\hat r/na_0$ times a harmonic polynomial, then the recurrence relation continues to have a reverse Bessel polynomial times a harmonic polynomial until the recursion ends because there are no more terms left in the harmonic polynomial. Hence, instead of starting with a power times a harmonic polynomial, we can easily compute the result if we start from a reverse Bessel polynomial times a harmonic polynomial. This allows us to immediately write down the following identity, which is illustrated graphically in Fig.~\ref{fig:bessel}:
\begin{align}
&\langle p_x,p_y,p_z|\theta_{m-l}\left (\frac{\hat r}{na_0}\right )P_h^l(\hat r_x,\hat r_y,\hat r_z)|\phi_{\lambda{=}n}\rangle\nonumber\\
&~~~~~~~=\frac{(na_0)^{2l}}{(i\hbar)^l}P_h^l(p_x,p_y,p_z)\langle p_x,p_y,p_z|\theta_{m}\left (\frac{\hat r}{na_0}\right )|\phi_{\lambda{=}n}\rangle.\nonumber\\
\end{align}

We already know how to complete the calculation of those expectation values. We simply use the definition of the reverse Bessel polynomial in Eq.~(\ref{eq: rev_bessel_poly_def}) and the results for the expectation value of powers of the radial coordinate in Eqs.~(\ref{eq: mu_final}) and (\ref{eq: q_final_sum}) to compute the final result. We find
\begin{align}
\langle p_x,p_y,p_z|&\theta_m\left (\frac{\hat r}{na_0}\right )|\phi_{\lambda{=}n}\rangle=\sum_{j=0}^m\frac{(m+j)!}{(m-j)!j!2^j}\langle p_x,p_y,p_z|\left (\frac{\hat r}{na_0}\right )^{m-j}
|\phi_{\lambda{=}n}\rangle\nonumber\\
&=\sum_{j=0}^m\frac{(m+j)!}{(m-j)!j!2^j}\frac{Q_{m-j}(\xi_n^2)}{\left [1+\left (\frac{na_0p}{\hbar}\right )^2\right ]^{m-j}}
\langle p_x,p_y,p_z
|\phi_{\lambda{=}n}\rangle\nonumber\\
&=\sum_{j=0}^m\frac{(m+j)!}{(m-j)!j!2^j}\frac{\sum_{k=0}^{\left \lfloor \frac{m-j+1}{2}\right \rfloor}(-1)^k\frac{(m-j+2)!(m-j+1)!}{2(m-j+1-2k)!(2k+1)!}\xi_n^{2k}}{\left [1+\left (\frac{na_0p}{\hbar}\right )^2\right ]^{m-j}}
\langle p_x,p_y,p_z
|\phi_{\lambda{=}n}\rangle
\end{align}
It turns out that these complicated summations can be simplified, but they require some significant computation. Because we have not seen the details of this elsewhere, we present them here for completeness.

The innermost summation can be schematically written as follows:
\begin{equation}
    \sum_{k=0}^{\left \lfloor \frac{n-1}{2}\right \rfloor}(-1)^k\left ( \begin{array}{c}
    n\\
    2k+1
    \end{array}\right )\xi_n^{2k},
\end{equation}
with $n=m-j+2$ and suppressing constant factors with respect to the summand index $k$. This summation can be easily done. Simply use the binomial theorem to re-express the difference of two powers as follows:
\begin{align}
    \frac{1}{2i\xi_n}\left [ (1+i\xi_n)^n-(1-i\xi_n)^n\right ] &=\sum_{k=0}^n\left ( \begin{array}{c}
    n\\
    k
    \end{array}\right )\left [ \frac{(i\xi_n)^k-(-i\xi_n)^k}{2i\xi_n}\right ]\nonumber\\
    &=\sum_{k=0}^{\left \lfloor \frac{n-1}{2}\right \rfloor}\left ( \begin{array}{c}
    n\\
    2k+1
    \end{array}\right )(-1)^k\xi_n^{2k},
\end{align}
because only terms with odd exponents contribute to the sum on the right hand side of the top line (we changed $k\to 2k+1$ in the second line). This completes the innermost summation. 

The outermost sum becomes
\begin{align}
    &\sum_{j=0}^m\frac{(m+j)!(m-j+1)}{j!2^j}
    \frac{(1+\xi_n^2)^{j-m}}{2\xi_n}{\rm Im}[(1+i\xi_n)^{m-j+2}]\nonumber\\
    &~~~~~~~={\rm Im}\Biggr \{\frac{m!(1+i\xi_n)^{m+2}}{2\xi_n(1+\xi_n^2)^m}\sum_{j=0}^m\left (\begin{array}{c}
    m+j\\
    j\end{array}\right )(m-j+1)\left (\frac{1-i\xi_n}{2}\right )^j\Biggr \}.
    \label{eq: outer_sum}
\end{align}

This remaining sum can be performed exactly. We go through the details next. We start by investigating the summation
\begin{equation}
    \sum_{j=0}^m \left (\begin{array}{c}
    m+j\\
    j
    \end{array}\right ) z^j.
    \label{eq: sum_identity}
\end{equation}
Since we can write the binomial expansion as
\begin{equation}
    (1+x)^{m+j}=\sum_{k=0}^{m+j}\left (\begin{array}{c}
    m+j\\
    k
    \end{array}\right )x^k,
\end{equation}
we see that the binomial coefficient in the summation in Eq.~(\ref{eq: sum_identity}) is the coefficient of the $x^j$ term, so that
\begin{equation}
    \sum_{j=0}^m \left (\begin{array}{c}
    m+j\\
    j
    \end{array}\right ) z^j=\frac{1}{2\pi i}\oint \sum_{j=0}^{m}\frac{(1+x)^{m+j}}{x^j}z^j\frac{dx}{x},
\end{equation}
where the contour for the integral encircles the origin and lies within a circle whose radius is smaller than $z/(1-z)$. Summing the series gives
\begin{equation}
     \sum_{j=0}^m \left (\begin{array}{c}
    m+j\\
    j
    \end{array}\right ) z^j=\frac{1}{2\pi i}\oint(1+x)^m\frac{1-\left (\frac{(1+x)z}{x}\right )^{m+1}}{x(1-z)-z}dx.
\end{equation}
The only pole comes from $x=0$ (which is an order $m+1$ pole) and we find
\begin{equation}
     \sum_{j=0}^m \left (\begin{array}{c}
    m+j\\
    j
    \end{array}\right ) z^j=-\frac{z^{m+1}}{m!}\frac{d^m}{dx^m}\left . \left (\frac{(1+x)^{2m+1}}{x(1-z)-z}\right )\right |_{x=0}.
\end{equation}
Some simple algebra with the operators $zd/dz$ applied to the summation yields
\begin{equation}
 \sum_{j=0}^m \left (\begin{array}{c}
    m+j\\
    j
    \end{array}\right ) (m+1-j)z^j=\frac{z^{m+2}}{m!}\frac{d^m}{dx^m}\left .\left (\frac{(1+x)^{2m+2}}{[(x(1-z)-z]^2}\right )\right |_{x=0}.
\end{equation}

Using this result, we find Eq.~(\ref{eq: outer_sum}) becomes
\begin{equation}
    {\rm Im}\left \{ \frac{(1+\xi_n^2)^2}{2^{m+1}\xi_n}\frac{d^m}{dx^m}\left .\left (\frac{(1+x)^{2m+2}}{[x(1+i\xi_n)-1+i\xi_n]^2}\right )\right |_{x=0}\right \}.
\end{equation}
Taking the imaginary part yields
\begin{equation}
    \frac{1}{2^m}\frac{d^m}{dx^m}\left .\left (\frac{(1+x)^{2m-1}(1-x)}{\left [1-\frac{2xv}{(1+x)^2}\right ]^2}\right )\right |_{x=0},
    \label{eq: deriv_ident}
\end{equation}
with $v=2/(1+\xi_n^2)$. Clearly, only the coefficient of the $x^m$ term in the parenthesis will contribute. To isolate this term, we first note that for $|2vx/(1+x)^2|<1$, which will hold for all $v$ as $x\to 0$, we have
\begin{equation}
    \frac{1}{\left [1-\frac{2vx}{(1+x)^2}\right ]^2}=\sum_{n=0}^\infty (n+1)\left ( \frac{2vx}{(1+x)^2}\right )^n.
\end{equation}
Combining this with the binomial theorem, which says
\begin{equation}
    (1+x)^{2m-2n-1}=\sum_{j=0}^{2m-2n-1}\left (\begin{array}{c}
    2m-2n-1\\
    j
    \end{array}\right )x^j,
\end{equation}
for $2m-2n-1\ge 0$, we find the following: (i) when $n<m$, the coefficient of $x^m$ is
\begin{equation}
    \frac{(2v)^n}{2^m}(n+1)\left [ \left (\begin{array}{c}
    2m-2n-1\\
    m-n\end{array}\right )-\left (\begin{array}{c}
    2m-2n-1\\
    m-n-1\end{array}\right )\right ]=0;
\end{equation}
(ii) when $n=m$, the coefficient of $x^m$ is the coefficient of $x^m$ in 
\begin{equation}
    \sum_{k=0}^\infty (-1)^kx^k(1-x)(m+1)\frac{(2vx)^m}{2^m},
\end{equation}
which implies $k=0$ with the coefficient being $(m+1)v^m$; and (iii) when $n>m$, there are no terms proportional to $x^m$ since the minimal power of $x$ that appears is $x^{m+1}$. Hence, we learn that Eq.~(\ref{eq: deriv_ident}) is equal to $(m+1)!v^m$. This simplification then allows us to conclude that
\begin{equation}
    \langle p_x,p_y,p_z|\theta_m\left ( \frac{\hat r}{na_0}\right )|\phi_{\lambda{=}n}\rangle=\frac{(m+1)!2^m}{(1+\xi_n^2)^m}\langle p_x,p_y,p_z|\phi_{\lambda{=}n}\rangle.
    \label{eq: theta_me}
\end{equation}
It is remarkable that one finds such a simple result for this complex expression!

We are now ready to work on the general case. We begin with the definition of the momentum-space wavefunction in Eq.~(\ref{eq: momentum_wf}). The first step is to recognize that the term $P_h^l(\hat r_x,\hat r_y,\hat r_z)/\hat r^l$ can be moved to the left, because it commutes with $\hat r$ and $\hat p_r$. Then, the work in Sections II and III showed that
\begin{equation}
    (-i)^{n-l-1}\hat B_r^\dagger(l)\cdots \hat B_r^\dagger(n-2)\hat r^{n-1}|\phi_{\lambda{=}n}\rangle\propto\frac{\sqrt{(n-l-1)!}}{\sqrt{n[(n+l)!]^3}}\left (\frac{2}{na_0}\right )^l\hat r^lL_{n-l-1}^{2l+1}\left (\frac{2\hat r}{na_0}\right )|\phi_{\lambda{=}n}\rangle,
\end{equation}
where the left hand side is not normalized, but the right hand side is.
We need to re-express this polynomial in $\hat r$ in terms of an expansion with respect to reverse Bessel polynomials. Following Ref.~\cite{bessel_poly_book}, we can expand a polynomial of the form $f(z)=\sum_{k=0}^nf_kz^k$ in terms of reverse Bessel polynomials via
\begin{equation}
    f(z)=\sum_{k=0}^nc_k\theta_k(z),
\end{equation}
where $c_k$ satisfies
\begin{equation}
    c_k=\sum_{j=0}^k\frac{(-1)^j}{2^j}\frac{(k+j)!}{j!(k-j)!}\left[ f_{k+j}-(k+j+1)f_{k+j+1}\right ],
\end{equation}
with $f_k=0$ for $k>n$. We need to expand the polynomial $f(\hat r/na_0)=L_{n-l-1}^{2l+1}(2\hat r/na_0)$, so that we have
\begin{equation}
    f_k=(-1)^k\frac{(n+l)!}{k!}\left(\begin{array}{c}
    n+l\\
    2l+k+1\end{array}\right )2^k,
\end{equation}
for $k\le n-l-1$.
We next compute
\begin{align}
    f_{k+j}-(k+j+1)f_{k+j+1}&=(-1)^{k+j}2^{k+j}\frac{(n+l)!}{(k+j)!}\nonumber\\
&\times    \left [ \left (\begin{array}{c}
    n+l\\
    2l+k+j+1\end{array}\right )+2\left (\begin{array}{c}
    n+l\\
    2l+k+j+2\end{array}\right )\right ]\nonumber\\
    &=(-1)^{k+j}2^{k+j}\frac{(n+l)!}{(k+j)!}\frac{(n+l)!(2n-k-j)}{(2l+k+j+2)!(n-l-k-j-1)!}.\nonumber\\
\end{align}
Using this result, we find that $c_k$ satisfies
\begin{equation}
    c_k=(-1)^k2^k[(n+l)!]^2\sum_{j=0}^{min(k,n-l-k-1)}\frac{(2n-k-j)}{j!(k-j)!(2l+k+j+2)!(n-l-k-j-1)!}.
\end{equation}
The sum can be performed by employing the well-known Van Der Monde convolution of binomial coefficients,
\begin{equation}
    \sum_{j=0}^n\left (\begin{array}{c}
    n\\
    j\end{array}\right )\left (\begin{array}{c}
    m\\
    k-j\end{array}\right )=\left (\begin{array}{c}
    n+m\\
    k\end{array}\right );
\end{equation}
we break up the summation into two terms---one corresponding to $2n-k$ and the other to $-j$. After summing the terms, we end up with
\begin{equation}
    c_k=(-1)^k2^k\frac{2n(n+l)!(n+l+k)!}{k!(n-l-k-1)!(2l+2k+2)!}.
\end{equation}
Putting this all together, we have
\begin{align}
&(-i)^{n-l-1}\langle p_x,p_y,p_z|\hat B_r^\dagger(l)\cdots \hat B_r^\dagger(n-2)\hat r^{n-l-1}P_h^l(\hat r_x,\hat r_y,\hat r_z)|\phi_{\lambda{=}n}\rangle=\frac{\sqrt{(n-l-1)!}}{\sqrt{n[(n+l)!]^3}}
\left (\frac{2}{na_0}\right )^l\nonumber\\
&\times\langle p_x,p_y,p_z|\sum_{k=0}^{n-l-1} (-1)^k2^k\frac{2n(n+l)!(n+l+k)!}{k!(n-l-k-1)!(2l+2k+2)!}\theta_k\left (\frac{\hat r}{na_0}\right )
 P_h^l(\hat r_x,\hat r_y,\hat r_z)|\phi_{\lambda{=}n}\rangle\nonumber\\
&=\frac{\sqrt{(n-l-1)!}}{\sqrt{n[(n+l)!]^3}}
(na_0)^l\frac{P_h^l(p_x,p_y,p_z)}{(i\hbar)^l}\sum_{k=0}^{n-l-1} (-1)^k2^{k+l}\nonumber\\
&\times\frac{2n(n+l)!(n+l+k)!}{k!(n-l-k-1)!(2l+2k+2)!}\langle p_x,p_y,p_z|\theta_{k+l}\left (\frac{\hat r}{na_0}\right )
|\phi_{\lambda{=}n}\rangle\nonumber\\
&=\frac{\sqrt{(n-l-1)!n}}{\sqrt{(n+l)!}}
(na_0)^l\frac{P_h^l(p_x,p_y,p_z)}{(i\hbar)^l}\sum_{k=0}^{n-l-1} (-1)^k2^{2k+2l+1}\nonumber\\
&\times\frac{(n+l+k)!(k+l+1)!}{k!(n-l-k-1)!(2l+2k+2)!}\frac{1}{(1+\xi^2)^{k+l}}\langle p_x,p_y,p_z
|\phi_{\lambda{=}n}\rangle.\nonumber\\
\end{align}
It turns out that this involves the properly normalized Gegenbauer polynomial, but it requires some additional work to show this.

The Gegenbauer polynomial satisfies
\begin{align}
    C_{n-l-1}^{l+1}\left (\frac{-1+\xi_n^2}{1+\xi_n^2}\right )
    &=\frac{(n+l)!}{(2l+1)!(n-l-1)!}\,_2F_1\left (-n+l+1,n+l+1,l+\frac{3}{2};\frac{1}{1+\xi_n^2}\right )\nonumber\\
    &=\frac{(n+l)!}{(2l+1)!(n-l-1)!}\sum_{k=0}^{n-l-1}(-1)^k\frac{(n-l-1)!}{k!(n-l-k-1)!}\frac{(n+l+k)!}{(n+l)!}\nonumber\\
    &~~~~~~\times\frac{1}{\prod_{j=1}^k\left (l+\frac{1}{2}+j\right )}\frac{1}{(1+\xi^2)^k}\nonumber\\
    &=\frac{1}{l!}\sum_{k=0}^{n-l-1}(-1)^k\frac{2^{2k+1}(n+l+k)!(k+l+1)!}{k!(n-l-k-1)!(2l+2k+2)!}\frac{1}{(1+\xi_n^2)^k},
\end{align}
which follows from using its representation in terms of a hypergeometric function $\,_2F_1$ and simplifying the expression to put it into a form similar to the one we have already generated. This gives
\begin{align}
\psi_{nl}(p_x,p_y,p_z)&=\frac{\sqrt{(n-l-1)!n}}{\sqrt{(n+l)!}}
(na_0)^l\frac{P_h^l(p_x,p_y,p_z)}{(i\hbar)^l} C_{n-l-1}^{l+1}\left (\frac{-1+\xi_n^2}{1+\xi_n^2}\right )\nonumber\\
&\times\frac{2^{2l}l!}{(1+\xi_n^2)^l}\langle p_x,p_y,p_z|\phi_{\lambda{=}n}\rangle.
\end{align}

The only remaining calculation is to find the normalized auxiliary Hamiltonian ground state wavefunction in momentum space. To do this, we use a translation operator to translate the momentum state at the origin to the state at $(p_x,p_y,p_z)$. Hence,
\begin{equation}
    \phi_{\lambda{=}n}(p_x,p_y,p_z)=\langle p_x,p_y,p_z|\phi_{\lambda{=}n}\rangle=\langle \vec{0}|e^{-\frac{i}{\hbar}\sum_\alpha p_\alpha \hat r_\alpha}|\phi_{\lambda{=}n}\rangle.
\end{equation}
Expanding in a power series, we have
\begin{equation}
    \phi_{\lambda{=}n}(p_x,p_y,p_z)=\sum_{m=0}^\infty \frac{(-1)^m}{m!}\left (\frac{i}{\hbar}\right )^m\langle \vec{0}|\left (\sum_\alpha p_\alpha \hat r_\alpha\right )^m|\phi_{\lambda{=}n}\rangle.
    \label{eq: mom_sum_zeroth}
\end{equation}
First we show that the matrix element vanishes when $m=1$. To see this, we use the property that the auxiliary Hamiltonian ground state is annihilated by $\hat A_\alpha(\lambda{=}n)$ for all $\alpha$. Then we directly compute
\begin{align}
\sum_\alpha p_\alpha\langle \vec{0}|\hat r_\alpha|\phi_{\lambda{=}n}\rangle&=\sum_\alpha p_\alpha \frac{na_0}{i\hbar}
\langle \vec{0}|\hat r \hat p_\alpha|\phi_{\lambda{=}n}\rangle=\sum_\alpha p_\alpha \frac{na_0}{i\hbar}\langle \vec{0}|[\hat r,\hat p_\alpha]|\phi_{\lambda{=}n}\rangle\nonumber\\
&=\sum_\alpha na_0p_\alpha \langle \vec{0}|\frac{\hat r_\alpha}{\hat r}|\phi_{\lambda{=}n}\rangle=\sum_\alpha \frac{(na_0)^2}{i\hbar}p_\alpha \langle \vec{0}|\hat p_\alpha |\phi_{\lambda{=}n}\rangle=0.
\end{align}
In general, we will show that all odd powers vanish. To see the general result, we consider the arbitrary case
\begin{align}
\sum_{\alpha_1\cdots\alpha_m}&p_{\alpha_1}\cdots p_{\alpha_m}\langle \vec{0}|\hat r_{\alpha_1}\cdots\hat r_{\alpha_m}|\phi_{\lambda{=}n}\rangle=
\sum_{\alpha_1\cdots\alpha_m}p_{\alpha_1}\cdots p_{\alpha_m}\frac{na_0}{i\hbar}\langle\vec{0}|\hat r_{\alpha_1}\cdots \hat r_{\alpha_{m-1}}\hat r \hat p_{\alpha_m}|\phi_{\lambda{=}n}\rangle
\nonumber\\
&=\sum_{\alpha_1\cdots\alpha_m}p_{\alpha_1}\cdots p_{\alpha_m}\frac{na_0}{i\hbar}\langle\vec{0}|[\hat r_{\alpha_1}\cdots \hat r_{\alpha_{m-1}}\hat r, \hat p_{\alpha_m}]|\phi_{\lambda{=}n}\rangle
\nonumber\\
&=\sum_{\alpha_1\cdots\alpha_{m-2}}p_{\alpha_1}\cdots p_{\alpha_{m-2}}p^2na_0(m-1)\langle\vec{0}|\hat r_{\alpha_1}\cdots \hat r_{\alpha_{m-2}}\hat r |\phi_{\lambda{=}n}\rangle\nonumber\\
&~~~~~+\sum_{\alpha_1\cdots\alpha_m}p_{\alpha_1}\cdots p_{\alpha_m}na_0\langle\vec{0}|\hat r_{\alpha_1}\cdots \hat r_{\alpha_{m-1}}\frac{\hat r_{\alpha_m}}{\hat r} |\phi_{\lambda{=}n}\rangle
\nonumber\\
&=\sum_{\alpha_1\cdots\alpha_{m-1}}p_{\alpha_1}\cdots p_{\alpha_{m-2}}\frac{(na_0p)^2(m-1)}{i\hbar}\langle\vec{0}|\hat r_{\alpha_1}\cdots \hat r_{\alpha_{m-1}}\hat p_{\alpha_{m-1}} |\phi_{\lambda{=}n}\rangle
\nonumber\\
&~~~~~+\sum_{\alpha_1\cdots\alpha_m}p_{\alpha_1}\cdots p_{\alpha_m}\frac{(na_0)^2}{i\hbar}\langle\vec{0}|\hat r_{\alpha_1}\cdots \hat r_{\alpha_{m-1}}\hat p_{\alpha_m} |\phi_{\lambda{=}n}\rangle
\nonumber\\
&=\sum_{\alpha_1\cdots\alpha_{m-1}}p_{\alpha_1}\cdots p_{\alpha_{m-2}}\frac{(na_0p)^2(m-1)}{i\hbar}\langle\vec{0}|[\hat r_{\alpha_1}\cdots \hat r_{\alpha_{m-1}},\hat p_{\alpha_{m-1}}] |\phi_{\lambda{=}n}\rangle
\nonumber\\
&~~~~~+\sum_{\alpha_1\cdots\alpha_m}p_{\alpha_1}\cdots p_{\alpha_m}\frac{(na_0)^2}{i\hbar}\langle\vec{0}|[\hat r_{\alpha_1}\cdots \hat r_{\alpha_{m-1}},\hat p_{\alpha_m}] |\phi_{\lambda{=}n}\rangle
\nonumber\\
&=(m-1)(m+2)(na_0p)^2\sum_{\alpha_1\cdots\alpha_{m-2}}p_{\alpha_1}\cdots p_{\alpha_{m-2}}\langle \vec{0}|\hat r_{\alpha_1}\cdots\hat r_{\alpha_{m-2}}|\phi_{\lambda{=}n}\rangle.
\end{align}
The last line follows from the fact that all terms are proportional to $(m-1)(na_0p)^2$: we get $m-2$ terms when the $\hat p_{\alpha_{m-1}}$ operator commutes with the $\hat r_{\alpha_j}$ terms with $1\le j\le m-2$; 3 more terms when it commutes with the $\hat r_{\alpha_{m-1}}$ operator; and then 1 more term from the last line. What this identity shows is that we can remove two operators at a time from the product, generating $(na_0 p)^2$ multiplied by numbers. If $m$ is odd, we end with the term that has $m=1$, which we already showed vanished. So all odd powers give zero. If $m$ is even, and is given by $m=2m'$, then we find the matrix element becomes
\begin{equation}
    \langle \vec{0}|\left (\sum_\alpha p_\alpha \hat r_\alpha\right )^{2m'}|\phi_{\lambda{=}n}\rangle=
    (2m')!(m'+1)(na_0p)^{2m'}\langle \vec{0}|\phi_{\lambda{=}n}\rangle.
\end{equation}
Substituting into the summation in Eq.~(\ref{eq: mom_sum_zeroth}) yields
\begin{equation}
\phi_{\lambda{=}n}(p_x,p_y,p_z)=\sum_{m'=0}^\infty (-1)^{m'}(m'+1)\left (\frac{na_0p}{\hbar}\right )^{2m'}\langle \vec{0}|\phi_{\lambda{=}n}\rangle.
\end{equation}
The sum is the derivative of the geometric series, so we finally obtain
\begin{equation}
\phi_{\lambda{=}n}(p_x,p_y,p_z)=\frac{1}{(1+\xi_n^2)^2}\langle \vec{0}|\phi_{\lambda{=}n}\rangle.
\end{equation}
Because the harmonic polynomial $\hat P_h^l(p_x,p_y,p_z)/p^l$ is normalized when integrated over the angular coordinates, we choose the normalization constant to satisfy
\begin{equation}
    1=|\langle \vec{0}|\phi_{\lambda{=}n}\rangle|^2\int_0^\infty dp p^2 \frac{1}{\left [1+\frac{(na_0p)^2}{\hbar^2}\right ]^4},
\end{equation}
or
\begin{equation}
    \langle \vec{0}|\phi_{\lambda{=}n}\rangle=\frac{4\sqrt{2}(na_0)^{\frac{3}{2}}}{\sqrt{\pi}\hbar^{\frac{3}{2}}}.
\end{equation}

The final momentum space wavefunction becomes
\begin{align}
\psi_{nl}(p_x,p_y,p_z)&=\frac{\sqrt{2(n-l-1)!}}{\sqrt{\pi(n+l)!}}
n^2\left (\frac{na_0p}{\hbar}\right )^l\frac{P_h^l(p_x,p_y,p_z)}{(ip)^l} C_{n-l-1}^{l+1}\left (\frac{-1+\xi_n^2}{1+\xi_n^2}\right )\nonumber\\
&~~~~~\times\frac{2^{2l+2}l!}{(1+\xi_n^2)^{l+2}}\left (\frac{a_0}{\hbar}\right )^{\frac{3}{2}},
\end{align}
which is the conventional result for the momentum-space wavefunction of Hydrogen.

We conclude this quite technical section with some thoughts. It is in many respects quite amazing that one can derive the momentum wavefunction algebraically, when none of the operators that appear in the operator expression of the wavefunction can be evaluated in terms of eigenvalues when they operate on momentum eigenstates. Nevertheless, using just the commutation relations and the fundamental properties of the auxiliary Hamiltonian ground states, we can completely construct them. In many respects, the hardest part of the challenge is rearranging the results to fit into the standard formulas for the polynomial representations used in the standard treatment of these wavefunctions. Yes the algebra is lengthy, but we find it remarkable that all of this information is encoded within the commutation relations themselves (and the existence of position and momentum-space eigenfunctions at the origin).

\section{The conventional confluent hypergeometric  equation approach}

In this section, we relax the condition of not using calculus, and focus on how one can derive the conventional differential equation that the wavefunctions satisfy. The approach will also be based heavily on operator algebra, but it follows closely the differential equation approach used to solve the Hydrogen atom in a Cartesian coordinate basis~\cite{ajp_cartesian}. Because of the additional complications one encounters in formulating  a differential equation in momentum space, we forgo examining that problem here.

The ansatz for deriving the differential equation is that the wavefunction can be written in the following factorized form
\begin{equation}
    |\psi_{\lambda l}\rangle=f_{\lambda l}(\hat r)P_h^l(\hat r_x,\hat r_y,\hat r_z)|\phi_\lambda\rangle,
\end{equation}
where $|\phi_\lambda\rangle$ is the auxiliary Hamiltonian ground state, which satisfies $\hat{\mathcal H}(\lambda)|\phi_\lambda\rangle=E(\lambda)|\phi_\lambda\rangle$. At this stage, we do not yet require the parameter $\lambda$ to be a positive integer. Our goal is to find a differential equation for $f_{\lambda l}$. This turns out to be rather straightforward, given the operator identities we have already developed. But before proceeding too far, we need to discuss one technical detail. We assume that $f_{\lambda l}(\hat r)$ can be expressed as a power-series expansion. In this case, it is straightforward to note that
\begin{equation}
[f_{\lambda l}(\hat r),\hat p_r]=i\hbar \frac{df_{\lambda l}(\hat r)}{d\hat r},
\end{equation}
where the operator derivative notation above simply means that if $f_{\lambda l}(\hat r)=\sum_{n=0}^\infty a_n\hat r^n$, then $df_{\lambda l}(\hat r)/d\hat r=\sum_{n=1}^\infty na_n\hat r^{n-1}$.

Now we simply operate $\hat{\mathcal H}(\lambda{=}1)$ onto $|\psi_{\lambda l}\rangle$ and force the equation to satisfy an eigenvalue/eigenvector relationship. Hence, writing the Hamiltonian in terms of the kinetic and potential-energy pieces $\hat{\mathcal H}(1)=\hat T+\hat V$, we find the potential term commutes with the functions of coordinate operators and we obtain
\begin{align}
    \hat{\mathcal H}(1)|\psi_{\lambda l}\rangle&=[\hat T,f_{\lambda l}(\hat r)]P_h^l(\hat r_x,\hat r_y,\hat r_z)|\phi_\lambda\rangle+f_{\lambda l}(\hat r)[\hat T,P_h^l(\hat r_x,\hat r_y,\hat r_z)]|\phi_\lambda\rangle\nonumber\\
    &+f_{\lambda l}(\hat r)P_h^l(\hat r_x,\hat r_y,\hat r_z)\left \{\hat {\mathcal H}(\lambda)+\frac{\lambda-1}{\lambda}\hat V(\hat r)\right \}|\phi_{\lambda l}\rangle.
\end{align}
We examine each term in turn. First, since $\hat T_\perp$ commutes with $f_{\lambda l}(\hat r)$, we have
\begin{align}
    [\hat T,f_{\lambda l}(\hat r)]&=\frac{1}{2m}\left \{\hat p_r[\hat p_r,f_{\lambda l}(\hat r)]+[\hat p_r,f_{\lambda l}(\hat r)]\hat p_r\right \}\nonumber\\
    &=-i\frac{\hbar}{2m}\left \{\hat p_r \frac{df_{\lambda l}(\hat r)}{d\hat r}+\frac{df_{\lambda l}(\hat r)}{d\hat r}\hat p_r\right \}\nonumber\\
    &=-\frac{\hbar^2}{2m}\frac{d^2f_{\lambda l}(\hat r)}{d\hat r^2}-i\frac{\hbar}{m}\frac{df_{\lambda l}(\hat r)}{d\hat r}\hat p_r.
\end{align}
This operator acts on $\hat P_h^l(\hat r_x,\hat r_y,\hat r_z)|\phi_\lambda\rangle$. Using the fact that $[\hat p_r,P_h^l(\hat r_x,\hat r_y,\hat r_z)]=$\\ $-i\hbar lP_h^l(\hat r_x,\hat r_y,\hat r_z)/\hat r$ and Eq.~(\ref{eq: pr_state}) give
\begin{align}
    [\hat T,f_{\lambda l}(\hat r)]P_h^l(\hat r_x,\hat r_y,\hat r_z)|\phi_\lambda\rangle&=-\frac{\hbar^2}{2m}\frac{d^2f_{\lambda l}(\hat r)}{d\hat r^2}P_h^l(\hat r_x,\hat r_y,\hat r_z)|\phi_\lambda\rangle\nonumber\\
    &-\left \{\frac{\hbar^2(l+1)}{m\hat r}-\frac{\hbar^2}{m\lambda a_0}\right \}\frac{df_{\lambda l}(\hat r)}{d\hat r}P_h^l(\hat r_x,\hat r_y,\hat r_z)|\phi_{\lambda}\rangle.
\end{align}
We use Eq.~(\ref{eq: tperp_com1}) to evaluate the second term, which yields
\begin{equation}
    f_{\lambda l}(\hat r)[\hat T,P_h^l(\hat r_x,\hat r_y,\hat r_z)]|\phi_\lambda\rangle=\frac{e^2l}{\lambda \hat r}f_{\lambda l}(\hat r)P_h^l(\hat r_x,\hat r_y,\hat r_z)|\phi_\lambda\rangle.
\end{equation}
The third term becomes
\begin{equation}
    \left \{E(\lambda)-\frac{e^2(\lambda-1)}{\lambda\hat r}\right \}f_{\lambda l}(\hat r)P_h^l(\hat r_x,\hat r_y,\hat r_z)|\phi_\lambda\rangle.
\end{equation}
Assembling these three terms gives
\begin{align}
\hat{\mathcal H}(1)|\psi_{\lambda l}\rangle&=E(\lambda)|\psi_{\lambda l}\rangle+\left \{ -\frac{\hbar^2}{2m}\frac{d^2f_{\lambda l}(\hat r)}{d\hat r^2}-\frac{\hbar^2(l+1)}{m\hat r}\frac{df_{\lambda l}(\hat r)}{d\hat r}\right .
\nonumber\\
&+\frac{e^2}{\lambda}\frac{df_{\lambda l}(\hat r)}{d\hat r}-\left .\frac{e^2(\lambda -l -1)}{\lambda \hat r}f_{\lambda l}(\hat r)\right \}P_h^l(\hat r_x,\hat r_y,\hat r_z)|\phi_\lambda\rangle.\nonumber\\
\end{align}
For the state $|\psi_{\lambda l}\rangle$ to be an eigenvector, we need the object in the curly brackets to vanish. It turns out this is the standard confluent hypergeometric equation for the Hydrogen atom's radial wavefunction, but to see this, we need to make the radial coordinate operator dimensionless. As before, we define $\hat \rho=2\hat r/(\lambda a_0)$. Then, one can immediately verify that the expression in the curly brackets becomes (after multiplying by $\hat \rho\lambda^2a_0/2e^2$ and evaluating against a coordinate eigenstate $|r_x,r_y,r_z\rangle$)
\begin{equation}
    \rho \frac{d^2 f_{\lambda l}(\rho)}{d\rho^2}+(2l+2-\rho)\frac{df_{\lambda l}(\rho)}{d\rho}+(\lambda-l-1)f_{\lambda l}(\rho)=0.
\end{equation}
This is the confluent hypergeometric equation for the Hydrogen atom. One can then solve this using the Frobenius method and reproduce the wavefunction we worked out using other techniques in Sec. II.

\section{Conclusions}

In this work, we have presented the solution of the Hydrogen atom using a Cartesian operator factorization approach. While the calculation is lengthy, it has a number of interesting aspects to it. First, it shows how one can solve quantum mechanics problems in terms of sums of {\it noncommuting} factorizations. Second, it illustrates how one can find the wavefunctions in coordinate space and in momentum space employing the same methodology (although the algebra for one is more complex). Third, it demonstrates that one can solve these problems without spherical harmonics and purely algebraically. We feel it is remarkable that one can achieve all of these goals for both the energy eigenvalues {\it and} the wavefunctions. 

This representation-independent formulation of operator mechanics can be applied more widely than just for Hydrogen. We have already employed it for determining spherical harmonics~\cite{weitzman} for the particle-in-a-box~\cite{jacoby} and for the simple-harmonic oscillator~\cite{rushka-ho}; it can also be applied to other problems that can be solved analytically.  Furthermore, operator mechanics provides a framework where one separates the determination of the energy eigenfunctions from the calculation of the wavefunctions in coordinate or momentum space; this representation-independent approach places quantum mechanics in a more elegant mathematical setting and completes the development initiated by Pauli, Dirac, and Schr\"odinger at the dawn of quantum mechanics.

\begin{acknowledgments}

We learned about the Cartesian factorization from Barton Zwiebach via the MIT 8.05x MOOC. He learned it from Roman Jackiw.
We thank Jolyon Bloomfield from MIT and Tianlalu from StackExchange for help in deriving the identity in Eq.~(\ref{eq: theta_me}). This work was supported by the National Science Foundation under grants numbered PHY-1620555 and PHY-1915130. J.K.F. was also supported at Georgetown University by the McDevitt bequest. X.L. was supported by the Global Science Graduate Course (GSGC) program of the University of Tokyo for the later parts of this work.

\end{acknowledgments}

\appendix

\setcounter{section}{0}

\section{Algebraic derivation of the ground-state wavefunction in the coordinate representation.}

In principle, we can go through a similar procedure as we did for deriving the momentum wavefunction that computes the coordinate wavefunction employing operators in Cartesian space. But the derivation becomes painful and torturous (and this paper has already a lot of long algebra in it). If, on the other hand, we derive the formulas employing the radial momentum operator to translate the radial coordinate, the calculation becomes much simpler. We provide a quick sketch of this approach here. 

The one ansatz we make is that this ground-state wavefunction is a function of $r$ only. Then a straightforward analysis with the radial momentum operator~\cite{rushka} shows that we can write the radial coordinate eigenstate as
\begin{equation}
    |r\rangle=e^{-\frac{ir}{\hbar}\left (\hat p_r-i\frac{\hbar}{\hat r}\right )}|r{=}0\rangle.
\end{equation}
Note that this is a highly nontrivial relation that could not be guessed. It follows from systematically changing coordinates from a Cartesian system to a spherical system and applying those coordinate changes to the conventional translation operator. Details are provided elsewhere~\cite{rushka}. The apparent singularity as $r\to 0$ is needed to cancel a similar singularity that arises from the radial momentum operator.

Using this result, we have immediately that
\begin{equation}
\psi_{10}(r)=\langle r|\phi_{\lambda{=}1}\rangle=
\langle r{=}0|e^{\frac{ir}{\hbar}\left (\hat p_r+i\frac{\hbar}{\hat r}\right )}|\phi_{\lambda{=}1}\rangle.
\end{equation}
Now, we expand in a power series
\begin{equation}
    \psi_{10}(r)=\sum_{n{=}0}^\infty\frac{1}{n!}\left (\frac{ir}{\hbar}\right )^n\langle r{=}0|\left (\hat p_r+i\frac{\hbar}{\hat r}\right )^n|\phi_{\lambda{=}1}\rangle
\end{equation}
and note that
\begin{equation}
    \left (\hat p_r+i\frac{\hbar}{\hat r}\right )|\phi_{\lambda{=}1}\rangle=i\hbar\left (\frac{1}{a_0}-\frac{1}{\hat r}+\frac{1}{\hat r}\right )|\phi_{\lambda{=}1}\rangle=\frac{i\hbar}{a_0}|\phi_{\lambda{=}1}\rangle.
\end{equation}
This simple result allows us to compute all powers immediately. Next, we perform the sum and immediately find that
\begin{equation}
    \psi_{10}(r)=e^{-\frac{r}{a_0}}\langle r{=}0|\phi_{\lambda{=}1}\rangle,
\end{equation}
which produces the correct wavefunction up to the normalization constant, that we already discussed in the main text.


\begin{thebibliography}{99}
\bibitem{pauli}
W. Pauli, ``\"Uber das Wasserstoffspektrum vom Standpunkt der neuen Quantenmechanik,'' Z. Phys. {\bf 36}, 336--363 (1926).
\bibitem{schroedinger_wavea}
E. Schr\"odinger, ``Quantisierung als Eigenwertproblem I.,'' Ann. Phys. (Leipzig) {\bf 384}, 361--376 (1926).
\bibitem{schroedinger_waveb}
E. Schr\"odinger, `Quantisierung als Eigenwertproblem II.,'' Ann. Phys. (Leipzig) {\bf 384}, 489--527 (1926).
\bibitem{schroedinger_factor1}
E. Schr\"odinger, ``A Method of Determining Quantum-Mechanical Eigenvalues and  Eigenfunctions,'' Proc. Roy. Irish Acad. Sec. A: Math. and Phys. Sci.  {\bf 46}, 9--16 (1940/1941).
\bibitem{schroedinger_factor2}
E. Schr\"odinger, ``Further Studies on Solving Eigenvalue Problems by Factorization,'' Proc. Roy. Irish Acad. Sec. A: Math. and Phys. Sci.  {\bf 46}, 183--206 (1940/1941).
\bibitem{infeld_hull}
L. Infeld and T. E. Hull, ``The Factorization Method,'' Rev. Mod. Phys. {\bf 23}, 21--68 (1951).
\bibitem{ajp_cartesian}
G. R. Fowles, ``Solution of the Schrödinger equation for the hydrogen atom in rectangular 283 coordinates,'' Am. J. Phys. {\bf 30}, 308--309 (1962).
\bibitem{stone}
M. H. Stone, ``Linear Transformations in Hilbert Space. III. Operational Methods and Group Theory.'' Proc. Nat. Acad. Sci. (USA) \textbf{16}, 172--175 (1930).
\bibitem{von-neumann}
J. von Neumann, ``Die Eindeutigkeit der Schr\"odingerschen Operatoren,'' Math. Ann. {\bf 104}, 570--578 (1931).
\bibitem{harris_loeb}
L. Harris and A. L. Loeb, {\it Introduction to Wave Mechanics} (McGraw-Hill, New York, 1963).
\bibitem{green}
H. S. Green, {\it  Matrix Mechanics} (P. Noordhoff Ltd., Gr\"onigen, 1965).
\bibitem{bohm}
A. Bohm, {\it Quantum Mechanics Foundations and Applications}, 3rd ed. (Springer-Verlag, Inc., New York, 1993).
\bibitem{ohanian}
H. C. Ohanian, {\it Principles of Quantum Mechanics} (Prentice Hall, New York, 1989).
\bibitem{delange_raab}
O. L. de Lange  and R. E. Raab {\it Operator Methods in Quantum Mechanics} (Oxford Science Series, Oxford, 1992).
\bibitem{hecht}
K. T. Hecht, {\it Quantum Mechanics} (Springer-Verlag, Inc., New York, 2000).
\bibitem{binney_skinner}
J. Binney and D. Skinner, {\it The Physics of Quantum Mechanics} (Oxford University Press, Oxford, 2014).
\bibitem{schwabl}
F. Schwabl, {\it  Quantum Mechanics}, 4th Ed. (Springer-Verlag, Inc., New York, 2007).
\bibitem{razavy}
M. Razavy, {\it Heisenberg's Quantum Mechanics} (World Scientific Publishing Co., Singapore, 2011).
\bibitem{judd}
B. R. Judd, {\it Angular Momentum Theory for Diatomic Molecules} (Academic Press, New York, 1975).
\bibitem{fock}
V. Fock, ``Zur Theorie des Wasserstoffatoms,''
Z. Phys. \textbf{98}, 145--154 (1935).
\bibitem{andrianov}
A. A. Andrianov, N. V. Borisov, and M. V. Ioffe, ``The factorization method and quantum systems with equivalent energy spectra,'' Phys. Lett. \textbf{105A}, 19--22 (1984).
\bibitem{rushka}
M. Rushka, M. A. Esrick, W. N. Mathews Jr., and J. K. Freericks, ``Converting translation operators into plane polar and spherical coordinates and their use in determining quantum-mechanical wavefunctions in a representation-independent fashion,''
J. Math. Phys. \textbf{62}, 072102 (2021).
\bibitem{dirac}
P. A. M. Dirac, Proc. Roy. Soc. London. Series A, Math. and Phys. {\bf 111}, 281 (1926).
\bibitem{kramers}
H. A. Kramers, {\it Quantum Mechanics} (North-Holland, Amsterdam, 1957).
\bibitem{brinkman}
H. C. Brinkman, {\it Applications of spinor invariants in atomic physics} (North-Holland, Amsterdam, 1956).
\bibitem{powell_crasemann}
J. L. Powell and B. Crasemann, {\it Quantum Mechanics} (Addison-Wesley, New York, 1961).
\bibitem{weinberg}
S. Weinberg, {\it Lectures on Quantum Mechanics} (Cambridge University Press, Cambridge, 2012).
\bibitem{avery}
J. S. Avery, ``Harmonic polynomials, hyperspherical harmonics, and atomic spectra,'' J. Comp and Appl. Math. {\bf 233}, 1366--1379 (2010).
\bibitem{schiff}
L. I. Schiff, {\it Quantum Mechanics} (McGraw-Hill, New York, 1949).
\bibitem{radial_momentum}
G. Paz, ``The non-self-adjointness of the radial momentum operator in n dimensions,'' J. Phys. A: Math. Gen. {\bf 35}, 3727 (2002).
\bibitem{bessel_poly_book}
E. Grosswald, {\it Bessel Polynomials}, Lecture Notes in Mathematics, ed. by A. Dold and B. Eckmann, Vol. 698 (Springer-Verlag, Inc. New York, 1978).
\bibitem{weitzman}
M. Weitzman and J. K. Freericks, ``Calculating spherical harmonics without derivatives,'' Cond. Mat. Phys. \textbf{21}, 33002 (2018).
\bibitem{jacoby}
J. A. Jacoby,  M. Curran, D. R. Wolf, and J. K. Freericks, ``Proving the existence of bound states for attractive potentials in one-dimension and two-dimensions without calculus,'' Eur. J. Phys. \textbf{40}, 045404 (2019).
\bibitem{rushka-ho}
M. Rushka and J. K. Freericks, ``A completely algebraic solution of the simple harmonic oscillator,'' Am. J. Phys. \textbf{88}, 976--985 (2020).

\end{thebibliography}

\end{paracol}

\end{document}